\documentclass{tlp}

\usepackage{aggr}

\allowdisplaybreaks[1]

\renewcommand{\tup}[1]{\overline{#1}}

\newcommand{\triv}[1]{triv(#1)}
\renewcommand{\ult}[1]{ult(#1)}
\newcommand{\bnd}[1]{bnd(#1)}

\newcommand{\CR}{\C{R}}
\newcommand{\ACD}{\appr{\CD}}

\newcommand{\cc}[1]{\ensuremath{\text{#1}}}

\newcommand{\qed}{\hfill\ensuremath{\Box}}

\newcommand{\lm}[1]{\lVert #1\rVert}

\DeclareMathOperator{\lub}{lub}
\DeclareMathOperator{\glb}{glb}

\DeclareMathOperator{\lmin}{min}
\DeclareMathOperator{\lmax}{max}

\newcommand{\ignore}[1]{}

\newcommand{\eval}[2]{\llbracket #1 \rrbracket_{#2}}

\newcommand{\struct}[1]{\langle #1 \rangle}

\newcommand{\leqp}{\leq_p}

\newcommand{\ST}[1]{{{\cal S}t_{#1}}}
\newcommand{\Alt}[1]{{A_{#1}}}
\newcommand{\ExSt}{ST}
\newcommand{\ust}[1]{St^{\uparrow}_{#1}}
\newcommand{\lst}[1]{St^{\downarrow}_{#1}}

\newtheorem{remark}{Remark}

\renewcommand{\CFM}{\wp_F}

\title{Well-founded and Stable Semantics of Logic Programs with
  Aggregates} 
\author{Nikolay Pelov, Marc Denecker, Maurice Bruynooghe}

\submitted{1 November 2005}
\revised{14 April 2006}
\accepted{4 May 2006}

\begin{document}

\maketitle


\begin{abstract}
  In this paper, we present a framework for the semantics and the
  computation of aggregates in the context of logic programming. In
  our study, an aggregate can be an arbitrary interpreted second order
  predicate or function. We define extensions of the Kripke-Kleene,
  the well-founded and the stable semantics for aggregate programs.
  The semantics is based on the concept of a three-valued {\em
    immediate consequence operator} of an aggregate program.  Such an
  operator {\em approximates} the standard two-valued immediate
  consequence operator of the program, and induces a unique
  Kripke-Kleene model, a unique well-founded model and a collection of
  stable models. We study different ways of defining such operators
  and thus obtain a framework of semantics, offering different
  trade-offs between {\em precision} and {\em tractability}. In
  particular, we investigate conditions on the operator that guarantee
  that the computation of the three types of semantics remains on the
  same level as for logic programs without aggregates. Other results
  show that, in practice, even efficient three-valued immediate
  consequence operators which are very low in the precision hierarchy,
  still provide optimal precision. 
\end{abstract}


\section{Introduction}

This paper is a study of the semantics of an extension of logic
programming with aggregates. Aggregates are specific second order
functions or predicates ranging over sets. Standard examples are the
minimum of a subset of a partially ordered domain, the cardinality of
a set, the sum, the product and the average of a finite set of
integers or reals, etc.  Aggregates play an important role in
different areas. They are used and studied extensively in the context
of databases (confer the {\em group-by } statement). They were
introduced in the context of logic programming as declarative variants
of the {\em set\_of} and {\em bag\_of} procedures
\cite{Kemp91-ILPS,Mumick90-VLDB}. Recently, they were introduced in
the context of two extensions of logic programming, Answer Set
Programming \cite{Simons02-AI} and Abductive Logic Programming
\cite{VanNuffelen00-NMR}. Aggregates commonly show up in human expert
knowledge and expressions of computational problems. For instance:
\begin{itemize}
\item the query for the {\em average} result of students for some
  exam;
\item the property that the capacity of a room should exceed the
{\em number} of  students attending the course that takes
place in that room;
\item the {\em cardinality constraint} that a lecturer should not
teach more than 6 courses;
\item the property or constraint that the {\em sum } of the
  capacities of available power generators in some electricity factory should
  exceed some given lower bound;
\end{itemize}
These examples show that aggregates are important to express many
forms of human expert knowledge and computational problems in a direct
and natural way. For this reason, aggregates likely will be part of
computational logics and the languages of future knowledge based
systems.

We will study the semantics of sets of rules of the form
\[ A \gets \varphi\] 
where $A$ is an atomic formula and $\varphi$ a logic expression
possibly containing aggregate formulas. Such rule sets are a core part
in logic programming and extensions such as abductive logic
programming and extended logic programming, the sub-logic of answer set
programming. Rule sets occur also as definitions of intensional
predicates in deductive databases and as inductive definitions in the
logic {\em ID-logic}, an extension of classical logic with
generalized, non-monotone inductive definitions \cite{Denecker00-CL}.
Thus, the results of our study can be applied in the context of all
these logics.

In the context of logic programming, several extensions with
aggregates were proposed for subclasses of the formalism that we
consider here, in particular for monotone aggregate programs
\cite{Mumick90-VLDB,Ross97-JCSS} and stratified aggregate programs
\cite{DellArmi03-IJCAI,Mumick90-VLDB}.  Our work extends such
proposals in two ways. First, we consider more general rule sets
with arbitrary recursion over aggregates. Second, we develop a
framework of semantics including extensions of the three main
semantics that have been used in logic programming:
Kripke-Kleene semantics (i.e., three-valued completion semantics)
\cite{Fitting85-JLP}, stable semantics \cite{Gelfond88-ICSLP} and the
well-founded semantics \cite{VanGelder91-JACM}.

The foundation of our work is the algebraic theory of approximating
operators \cite{Denecker00-LBAI,Denecker04-IC}. Approximation theory
is a fixpoint theory of non-monotone lattice operators.  With any
lattice operator $O\colon L\to L$, it associates a family of approximating
operators $A\colon L^2\to L^2$ on the product lattice $L^2$.  The fixpoint
theory associates with every approximating operator $A$ different
types of fixpoints: a Kripke-Kleene fixpoint and a well-founded
fixpoint, both in the bilattice $L^2$ and a set of $A$-stable
fixpoints of $O$ in the lattice $L$. In \cite{Denecker00-LBAI} it was
shown that the three-valued Fitting operator $\Phi_P$
\cite{Fitting85-JLP} is an approximation of the immediate consequence
operator $T_P$ of a logic program $P$ and that the different types of
fixpoints of $\Phi_P$ corresponds to the Kripke-Kleene, the
well-founded and the stable models of $P$. 

In \cite{Denecker04-IC}, the class of approximations of $O$ was
further investigated. The collection of approximations of a lattice
operator $O$ is ordered by a precision order. More precise
approximations have a more precise Kripke-Kleene and well-founded
fixpoint, and have more stable fixpoints. It was shown that $O$ has a
most precise approximation, called the ultimate approximation of $O$
which has the most precise semantics.

 
In the context of logic programming, approximation theory induces a
family of Kripke-Kleene, a family of well-founded and a family of
stable semantics, generated by the class of approximations of the
immediate consequence operator $T_P$. Basically, each family
formalizes similar intuitions but in different degrees of
precision. In \cite{Denecker04-IC}, the ultimate and the standard
versions of these semantics are investigated. It follows from the
general theory that the ultimate versions of the semantics are more
precise than the standard semantics.  Also, ultimate semantics have
elegant semantic properties which do not always hold for the standard
semantics based on the Fitting operator. For instance, substituting a
rule body $B$ by a formula $B'$ which is equivalent with $B$ in
classical logic, is equivalence preserving\footnote{In the sense of
  having the same set of models.} in the ultimate semantics 
but not in the standard semantics. Also, the ultimate well-founded
model of a program with a monotone $T_P$ is the least fixpoint of
$T_P$. On the negative side, applying the ultimate approximation is
computationally harder, and it was shown that computing the three
types of ultimate semantics for propositional programs is one level
higher in the polynomial hierarchy than the standard versions of the
same semantics. It was also shown that for important
classes of logic programs, standard and ultimate semantics coincide.
In fact, it seems that both semantics only differ for programs
containing a recursively defined predicate $p$ whose definition
also uses reasoning by cases of the form $(p \land\dots) \lor (\lnot p \land \dots)$.
In practice, such programs seem to be rare (we are unaware of any
practical program with this feature). Thus, the standard semantics
based on the Fitting operator and the ultimate semantics based on the
ultimate approximation are two very close points in the hierarchy of
semantics induced by approximation theory and represent different
trade-offs between precision and complexity.

In this work we apply the approximation theory in the context of rule
sets with aggregate expressions. We extend the two-valued immediate
consequence operator $T_P$ for aggregate programs, define several
different approximating operators of it and study the semantics
obtained from them. One operator is the ultimate approximation of
$T_P$. The three types of ultimate semantics obtained from this
operator extend the corresponding ultimate semantics for logic
programs. They also have the same attractive semantical properties and
the high computational complexity. So, we also study extensions of the
standard Kripke-Kleene, well-founded and stable semantics of logic
programs.  To achieve this, we propose the concept of a three-valued
aggregate relation approximating a given aggregate relation. We use
this concept to define an extension of the Fitting operator $\Phi_P$ to
the case of programs with aggregates.  Since an aggregate relation is
approximated by a class of three-valued aggregate relations, we obtain a
sub-family of approximations of $T_P$, all of which coincide with the
Fitting operator $\Phi_P$ in case $P$ does not contain aggregate
expressions. Just as in the case of logic programming without
aggregates, the different semantics based on the different
approximation operators represent close points in the hierarchy of
semantics induced by approximation theory and provide different
trade-offs between precision and complexity.


\section{Fixpoint Theory of Monotone and Non-monotone Operators}
\label{sec:at}

We now present the necessary background on Approximation Theory. For
more information we refer to \cite{Denecker04-IC}.

A structure $\struct{L,\leq}$ is a {\em poset} if $\leq$ is a reflexive,
asymmetric and transitive binary relation on $L$. A poset
$\struct{L,\leq}$ is a {\em chain} if $\leq$ is a total order, i.e., for each
$x, y \in L$, either $x\leq y$ or $y\leq x$. Sometimes, when the order
relation $\leq$ is clear from the context we denote a poset simply with
its domain $L$.

A poset $\struct{L,\leq}$ is {\em chain-complete} if each chain $S\subseteq
L$ has a least upper bound $\lub(S)$ in $L$. Since the empty set is a
chain, a chain-complete poset has a least element $\bot$.

A poset $\struct{L,\leq}$ is a {\em complete lattice} if each subset $S$
of $L$ has a least upper bound $\lub(S)$ and a greatest lower bound
$\glb(S)$ in $L$. In particular, $L$ has a least element $\bot$ and a
largest element $\top$. A complete lattice is chain-complete.

An operator $O\colon L\to L$ on a poset $\struct{L,\leq}$ is
$\leq$-monotone if for each $x, y\in L$, $x\leq y$ implies $O(x)\leq
O(y)$. A monotone operator $O$ on a chain-complete poset
$\struct{L,\leq}$ has a {\em least fixpoint} $\lfp(O)$.
This fixpoint can be constructively 
computed as a sequence of {\em powers} of $O$ defined as follows:
\begin{align*}
  O\uparrow^0(x) & = x \\
  O\uparrow^{\alpha+1}(x) &= O(O\uparrow^\alpha(x)) \\
  O\uparrow^\lambda(x) &= \lub(\{O\uparrow^\alpha(x)\vbar \alpha<\lambda\}) \text{ for a limit ordinal } \lambda
\end{align*}

\begin{proposition} \label{prop:fp}
  If $O\colon L\to L$ is a monotone operator and $L$ is a chain-complete poset
  then there exists an ordinal $\alpha$ such that $O\uparrow^\alpha(\bot)=\lfp(O)$. The
  least such ordinal is called the {\em closure ordinal} of $O$ and is
  denoted by $\infty$.
\end{proposition}

Approximation theory is an extension of the above fixpoint theory to
the case of arbitrary (non-monotone) lattice operators $O$. The
following basic concepts are needed.

Given a complete lattice  $\struct{L,\leq}$, its {\em bilattice} is
the structure $\struct{L^2,\leq, \leqp}$ where for all $x, y, x', y'
\in L$, 
\[ \begin{array}{ll}
  (x,y) \leq (x',y') &\text{ if and only if }
  x \leq x' \text{ and } y \leq y' \\
  (x,y) \leq_p (x',y')& \text{ if and only if }
  x \leq x' \text{ and } y' \leq y
\end{array} \]
The order $\leq$ on $L^2$ is called the {\em product order}, while
$\leqp$ is called the {\em precision order}. Both orders are complete
lattice orders on $L^2$.  We are interested only in a subset of $L^2$.
A pair $(x,y)$ is {\em consistent} if $x\leq y$ and {\em exact} if $x=y$.
The set of consistent pairs is denoted by $L^c$.

A basic intuition in approximation theory is that a consistent pair
$(x,y)$ approximates an element by a lower and an upper bound.  Hence,
$(x,y)$ {\em approximates} any element in the interval $[x,y]=\{z\in
L\vbar x\leq z\leq y\}$. More precise pairs approximate fewer elements:
$(x,y)\leq_p(x_1,y_1)$ implies $[x,y]\supseteq[x_1,y_1]$.  Pairs
$(x,x)$ are called exact because they approximate a single element
$x$. The set of exact pairs represents the embedding of $L$ in $L^c$.
The product order is a complete lattice order on $L^c$ while $\leqp$
is chain-complete on $L^c$.  Hence, any $\leq$-monotone or
$\leqp$-monotone operator on $L^c$ has a least fixpoint in the $\leq$ or
$\leqp$ order. Notice also that $(\bot,\top)$ is the $\leqp$-least element
of $L^c$ while the $\leqp$-maximal elements of $L^c$ are precisely the
set of exact elements.
\ignore{ In case $\struct{L,\leq}$ is a lattice, then
  $\struct{L^2,\leq, \leqp}$ is called the bilattice. Both $\leq$ and
  $\leqp$ are complete lattice orders on $L^2$.  The product order is
  a complete lattice order on $L^c$ while $\leqp$ is chain-complete on
  $L^c$.  Hence, any $\leq$-monotone or $\leqp$-monotone operator on
  $L^c$ has a least fixpoint. Notice also that $(\bot,\top)$ is the
  $\leqp$-least element of $L^c$ while the $\leqp$-maximal elements of
  $L^c$ are exactly the set of exact elements.}

\begin{example} \label{ex:three}
  Consider the lattice $\C{TWO}=\{\false,\true\}$ of classical truth
  values ordered as $\false < \true$. We denote the set $\C{TWO}^c$ of
  consistent approximations of $\C{TWO}$ with $\C{THREE}$. 

  The set $\C{THREE}$ corresponds to the set of truth values
  $\{\appr{\false},\appr{\undef},\appr{\true}\}$ used in three-valued logic.  The exact
  pairs $(\false,\false)$ and $(\true,\true)$, called {\em false} and
  {\em true} correspond to the values $\appr{\false}$ and $\appr{\true}$, while
  $(\false,\true)$ corresponds to $\appr{\undef}$, called {\em undefined}.
  The product order on $\C{THREE}$ corresponds to the truth order
  $\appr{\false} < \appr{\undef} <\appr{\true}$. The precision order on $\C{THREE}$
  corresponds to the order $\appr{\undef} <_p \appr{\false}, \appr{\undef} <_p \appr{\true}$, and
  is sometimes called the knowledge order.
  
  We can define logical connectives in $\C{THREE}$ in the following
  way. Conjunction $\land$ and disjunction $\lor$ of two elements are defined
  as the greatest lower bound and the least upper bound with respect
  to the product order $\leq$. The negation operator is defined as
  $\lnot(x,y)=(\lnot y,\lnot x)$. In particular, $\lnot\appr{\false}=\appr{\true}$, $\lnot\appr{\true}=\appr{\false}$,
  and $\lnot\appr{\undef}=\appr{\undef}$. The truth tables of the connectives $\land$, $\lor$,
  and $\lnot$ are the same as in Kleene's strong three-valued logic. \qed
\end{example}

\begin{definition}[Approximating Operator] \label{def:ap}
  Let $O\colon L\to L$ be an operator on a complete lattice $\langle L,\leq\rangle$. We say that
  $A\colon L^c\to L^c$ is an {\em approximating operator} of $O$ if the
  following conditions are satisfied:
  \begin{itemize}
  \item $A$ {\em extends} $O$, i.e., $A(x,x)=(O(x),O(x))$ for every
    $x\in L$;
  \item $A$ is $\leq_p$-monotone.
  \end{itemize}
\end{definition}
We denote the projections of an approximating operator $A\colon L^c\to
L^c$ on the first and second components with $A^1$ and $A^2$, i.e., if
$A(x,y)=(u,v)$ then $A^1(x,y)=u$ and $A^2(x,y)=v$.

There is a simple and natural intuition behind the concept of an
approximating operator $A$. Any tuple $(x,y)\in L^c$ can be viewed as
a non-empty interval $[x,y]=\{z\vbar x\leq z \leq y\}$. It is easy to see
that for any $z\in [x,y]$, $A(x,y)\leq_p A(z,z) = (O(z),O(z))$ which
means that the set $O([x,y])=\{O(z)\vbar z\in [x,y]\}$ is a subset of the
interval $A(x,y)$. 
Hence, $A^1(x,y)$ is a lower estimate and $A^2(x,y)$ is an upper
estimate to $O(z)$, for each $z$ in $[x,y]$.

The $\leq_p$-monotonicity of $A$ guarantees that $A$ has a least
fixpoint called the {\em Kripke-Kleene} fixpoint of $A$ and denoted
by $KK(A)$. This fixpoint approximates all fixpoints of $O$, i.e.,
if $x=O(x)$ then $KK(A)\leq_p(x,x)$.

Next we define the stable and well-founded fixpoints of $A$.  With a
fixed element $b\in L$, we can associate an operator $A^1(\cdot,b)$
mapping any $x\in[\bot,b]$ to $A^1(x,b)$. The operator $A^1(\cdot,b)$
is monotone but as a function from $[\bot,b]$ to $L$, in general it is
not internal in $[\bot,b]$. Similarly, for a fixed element $a\in L$,
the operator $A^2(a,\cdot)\colon [a,\top] \to L$ is monotone, but in
general is not internal in $[a,\top]$.

\begin{definition}
  Let $\langle L,\leq\rangle$ be a complete lattice.
  The {\em upper stable operator} $\ust{A}\colon L\to L$ is defined as
  \[ \ust{A}(a) = glb(\{ x \in [a,\top] \colon A^2(a,x) \leq x\}) \]
  and the {\em lower stable operator} $\lst{A}\colon L\to L$ is defined as:
  \[ \lst{A}(b) = glb(\{ x \in [\bot,b] \colon A^1(x,b) \leq x\}). \]
\end{definition} 
The upper stable operator maps a lattice element $a$ to the greatest
lower bound of the set of pre-fixpoints of $A^2(a,\cdot)$. If $A^2(a,\cdot)$
is internal in $[a,\top]$ then due to its monotonicity,
$\ust{A}(a)$ is its least fixpoint. The lower stable operator maps a
lattice element $b$ to the greatest lower bound of the set of
pre-fixpoints of $A^1(\cdot,b)$. This set may be empty, in which case
$\lst{A}(b)=\top$. However, if $A^1(\cdot,b)$ is internal in
$[\bot,b]$, then $\lst{A}(b)$ is its least fixpoint.

\begin{definition} \label{def-stable-operator}
  The {\em stable revision operator} $\ST{A}\colon L^2\to L^2$ where $\langle L,\leq\rangle$
  is a complete lattice is defined as follows:
  \[ \ST{A}(a,b) = (\lst{A}(b),\ust{A}(a)). \]
\end{definition}

In general, the stable revision operator is not internal in the set
$L^c$. However, there is a subclass of $L^c$ on which this operator
has very nice properties. It is defined as the intersection of the
following subclasses: 
\begin{itemize}
\item A pair $(a,b) \in L^c$ is {\em $A$-reliable}, if $(a,b) \leq_p
  A(a,b)$.
\item A pair $(a,b) \in L^c$ is {\em $A$-prudent}, if $a \leq \lst{A}(b)$.
\end{itemize}
It is easy to see that if $(a,b)$ is $A$-reliable, then the operators
$A^1(\cdot,b)$ and $A^2(a,\cdot)$ are internal in their domain. On the
other hand, if $(a,b)$ is $A$-prudent, we can guarantee that $a$ is a
safe underestimate of all fixpoints below $b$ of the operator $O$.

Intuitively, the stable revision operator implements two quite
different approximation processes, one to refine the upper estimate
$b$ and one to refine the lower estimate $a$. Given a current upper
estimate $b$, we compute a new lower estimate by an iterative process
$x_0=\bot, x_1=A^1(x_0,b), \dots, x_{i+1}=A^1(x_i,b), \dots$ until a
fixpoint is reached. In each stage, we use $A$ to approximate
$O([x_i,b])$ from below, i.e., by setting $x_{i+1} := A^1(x_i,b) \leq
O(z)$, for each $z\in [x_i,b]$. It is easy to see that each computed
$x_i$ is a lower estimate to each fixpoint of $O$ below $b$, and the
limit $\lst{A}(b)$ is the best lower bound we can obtain through $A$
to the set of these fixpoints.  On the other hand, the refined upper
estimate is computed as a limit of the sequence $y_0=a,
y_1=A^2(a,y_0), \dots, y_{i+1}=A^2(a,y_i), \dots$.  The goal is to
eliminate non-minimal, {\em non-reachable} fixpoints of $O$ above the
current lower estimate $a$.  Assuming that $a \leq O(a) (= A^2(a,a) =
y_1)$, all points in $[a,O(a)]$ are considered reachable.  On the next
level, also points in $O([a,O(a)])$ above $a$ are of interest, and we
can approximate these points from above by computing $A^2(a,y_1)=y_2$.
This process is continued until the fixpoint $\ust{A}(a)$ is reached
and this fixpoint is taken as the new upper bound.

In \cite{Denecker04-IC}, it was shown that the set $L^{rp}$ of pairs
that are $A$-reliable and $A$-prudent contains $(\bot,\top)$, is
chain-complete, and the stable revision operator is an internal,
$\leq_p$-monotone operator in $L^{rp}$. It follows that this operator
has a least fixpoint, called the {\em well-founded} fixpoint of $A$
and denoted by $WF(A)$. All consistent fixpoints of $\ST{A}$ are
$A$-reliable and $A$-prudent. They are called {\em stable
  fixpoints} of $A$ and they are $\leq$-minimal fixpoints of $A$.
The subset of exact stable fixpoints is denoted by $\ExSt(A)$.  Exact
stable fixpoints can be characterized alternatively as follows: $x\in
L$ is an exact stable fixpoint if and only if $O(x)=x$ and
$\lfp(A^1(\cdot,x))=x$. We have the following lemma.
\begin{lemma}\label{lem:mf-mm}
  A stable fixpoint of $A$ is a minimal pre-fixpoint of $O$.
\end{lemma}
The inverse however is not true: not every minimal pre-fixpoint of $O$
is a stable fixpoint of $A$.
\begin{proof}
  Let $x$ be a stable fixpoint of $A$ and $y$ a pre-fixpoint of $O$
  such that $y\leq x$. By anti-monotonicity of $A^1$ in the second
  argument, it holds that $A^1(y,x) \leq A^1(y,y) = O(y) \leq y$.
  Hence, $y$ is a pre-fixpoint of $A^1(\cdot,x)$ and since $x$ is the
  least fixpoint of this monotone operator, this implies that $x\leq
  y$.
\end{proof}

\ignore{
By Proposition~\ref{prop:fp}, the well-founded fixpoint of $A$ which
is the least fixpoint of $\ST{A}$ can be computed as
$\ST{A}\uparrow\infty(\bot,\top)$.  If we look closer at this sequence
we notice that there are two interleaved but independent threads of
computations. If we look at two consecutive steps of the $\ST{A}$
operator:
\begin{align*}
  \ST{A}(x_1,y_1) & = (\lfp(A^1(\cdot,y_1)),\lfp(A^2(x_1,\cdot))) = (x_2,y_2) \\
  \ST{A}(x_2,y_2) & = (\lfp(A^1(\cdot,y_2)),\lfp(A^2(x_2,\cdot))) = (x_3,y_3)
\end{align*}
we can see that the value of $x_3$ does not depend on $y_1$ as in the
computation below:
\begin{align*}
  \ST{A}(x_1,\_) & = (\_,\lfp(A^2(x_1,\cdot))) = (\_,y_2) \\
  \ST{A}(\_,y_2) & = (\lfp(A^1(\cdot,y_2)),\_) = (x_3,\_).
\end{align*}
which can be further simplified:
\begin{align*}
  \ST{A}^2(x_1) & = \lfp(A^2(x_1,\cdot)) = y_2 \\
  \ST{A}^1(y_2) & = \lfp(A^1(\cdot,y_2)) = x_3.
\end{align*}
To model this process we define the operator
\[ \Alt{A}(x) = \lfp(A^1(\cdot,\lfp(A^2(x,\cdot)))). \]
The operator $\Alt{A}$ is monotone and its least fixpoint is equal to
the under-estimate of the well-founded fixpoint $WF(A)=(u,v)$, i.e.,
$\lfp(\Alt{A})=u$. The over-estimate $v$ of the $WF(A)$ can be
computed as $v=\lfp(A^2(u,\cdot))$. The operator is closely related to the
alternating fixpoint computation of the well-founded model
\cite{VanGelder93-JCSS}.

}

In general, a lattice operator $O\colon L\to L$ may have many approximating
operators. For any pair $A$, $B$ of approximations of $O$, we define
$A\leqp B$ if and only if for each $(x,y)\in L^c$, $A(x,y)\leqp B(x,y)$.
The following result about the relationship between the different
classes of fixpoints of $A$ and $B$ was proven in \cite{Denecker04-IC}.

\begin{theorem}\label{TPrec}
  If $A\leqp B$, then $KK(A) \leqp KK(B)$, $WF(A) \leqp WF(B)$ and
  $\ExSt(A) \subseteq \ExSt(B)$.
\end{theorem}

So, more precise approximating operators lead to more precise
Kripke-Kleene and Well-founded fixpoints, and to more exact stable
fixpoints. It turns out that $O$ has a most precise approximation
$U_O$ called the {\em ultimate approximation} of $O$. It is defined
as:
\[ U_O(x,y) = (\glb(O([x,y])),\lub(O([x,y]))) \]
where $O([x,y]) = \{O(z) \vbar z \in [x,y]\}$. 
The Kripke-Kleene, stable and well-founded fixpoints of $U_O$ are
called the {\em ultimate Kripke-Kleene, ultimate stable and ultimate
  well-founded fixpoints of $O$}.

\begin{theorem}[\citeNP{Denecker04-IC}]
  The ultimate Kripke-Kleene and ultimate well-founded fixpoint of $O$
  is the most precise of all Kripke-Kleene and well-founded fixpoints
  of all approximations of $O$. The set of ultimate exact stable
  fixpoints includes all exact stable fixpoints of all approximations
  $A$ of $O$.
\end{theorem}

A special case arises when $O$ is monotone.

\begin{theorem}[\citeNP{Denecker04-IC}]\label{TUltMono}
  If $O$ is monotone, then for every $(x,y)\in L^c$,
  $U_O(x,y)=(O(x),O(y))$ and its ultimate well-founded fixpoint is
  exact and is the least fixpoint of $O$.
\end{theorem}

In \cite{Denecker00-LBAI}, it was shown that the Kripke-Kleene, the
well-founded and the stable semantics of a logic program $P$
correspond to Kripke-Kleene, well-founded and exact stable fixpoints
of the three-valued immediate consequence operator $\Phi_P$ of $P$
defined by \citeN{Fitting85-JLP}. In \cite{Denecker03-AI}, analogous
results were obtained in the context of default and autoepistemic
logic. This shows that approximation theory formalizes an important
non-monotonic principle.

\section{Aggregates} \label{sec:aggr}

\subsection{Aggregate Functions and Relations} \label{Sec-aggr}

In this text, an aggregate is understood as a second-order $n$-ary
function or relation having at least one set argument. For simplicity,
we assume that $n=1$ in case of aggregate functions and $n=2$ in case
of aggregate relations. We denote the power-set of a set $D$
with $\PS{D}$.

\begin{definition}[Aggregate Functions and Relations] \label{def:ar}
  Let $D_1$ and $D_2$ be domains.  An {\em aggregate function} is any
  function $\F\colon\PS{D_1}\to D_2$.  An {\em aggregate relation} is
  any relation $\R\subseteq\PS{D_1}\times D_2$.
\end{definition}

We use $\F$ to denote an aggregate function and $\R$ to denote an
aggregate relation. Although many aggregates are functions, for
uniformity and convenience sake, our theory below is developed for
aggregate relations. If an aggregate function $\F$ is used in a
context which requires an aggregate relation, $\F$ is understood as
its {\em graph} $\Aggr{G}_{\F}$ which is the aggregate relation
defined as $\Aggr{G}_{\F}=\{ (S,d) \vbar \F(S)=d \}$.

We now define a number of standard aggregate functions and relations
which we study in this paper. We start with aggregate relations in the
context of a poset $\langle D,\leq\rangle$:
\begin{definition} \label{Def-standard-aggr-relations}
  \begin{itemize}
  \item $\Glb\subseteq\PS{D}\times D$ -  defined as $\{(S,d) \vbar
  S\in \PS{D}  \text{ and }d =\Glb(S) \}$.
  \item $\Lub\subseteq\PS{D}\times D$ -  defined as $\{(S,d) \vbar
  S\in \PS{D}  \text{ and }d =\Lub(S) \}$.

 \item  $\Lb\subseteq\PS{D}\times D$ - defined as $\{(S,d) \vbar
    \forall x\in S, d\leq x\}$.
 \item  $\Ub\subseteq\PS{D}\times D$ - defined as $\{(S,d) \vbar
    \forall x\in S, x\leq d\}$.
 \item  $\Min \subseteq\PS{D}\times D$ - defined as $\{(S,d) \vbar  d
    \text{ is a  minimal element of } S\}$.
 \item  $\Max \subseteq\PS{D}\times D$ - defined as $\{(S,d) \vbar  d
    \text{ is a  maximal element of } S\}$.
  \end{itemize}
\end{definition}

The aggregate relations $\Glb$ and $\Lub$ are (graphs of) partial
aggregate functions. If $\langle D,\leq\rangle$ is a complete lattice
then $\Glb$ and $\Lub$ are (graphs of) total aggregate functions. If
$\langle D,\leq\rangle$ is a totally ordered set then $\Min$ and
$\Max$ represent partial functions. If in addition $D$ is
finite then $\Min$ and $\Max$ represent total functions and
$\Min=\Glb$ and $\Max=\Lub$.

Next, we define aggregate functions on finite sets of numbers. Below,
we assume that $D$ is an arbitrary domain and $D'$ is a Cartesian
product $D_1\times \dots \times D_n$ in which $D_1$ is the set of real
numbers $\setR$. Also, we denote the set of all finite subsets of a
domain $D$ by $\CFM(D)$.

\begin{definition} \label{def:aaf}
  \begin{itemize}
  \item $\Count\colon \CFM(D)\to\setN$ defined as $\displaystyle\Count(S)= \size{S}$, the cardinality of $S$;
  \item $\Sum\colon \CFM(D')\to\setR$ defined as
    $\displaystyle\Sum(S)=\hspace{-1em}\sum_{(x_1,\ldots,x_n)\in S}\hspace{-1em}x_1$;
  \item $\Prod\colon \CFM(D')\to\setR$ defined as
    $\displaystyle\Prod(S)=\hspace{-1em}\prod_{(x_1,\ldots,x_n)\in S}\hspace{-1em}x_1$;
  \item $\Avg\subseteq \CFM(D')\times\setR$ (Average) - the graph of a partial aggregate
    function defined only for non-empty sets as $(S,d)\in\Avg$ if
    $d=\Sum(S)/\Count(S)$.
  \end{itemize}
\end{definition}

In the definition of $\Sum, \Prod$ and $\Avg$, only the first element
of a tuple is used to compute the value. The reason to introduce the
other arguments is to be able to count one number multiple times.
That is, a set $S \subseteq \setR \times D_2\times \dots \times D_n$
represents a {\em multiset} of real numbers. For example, when
counting the total capacity of a building consisting of different
rooms, we need to count the capacity of a room as many times as
there are rooms with that capacity.



All these aggregate functions have no natural extensions to infinite
sets. However, their graphs $\Aggr{G}_\F$ can be considered as
aggregate relations on arbitrary sets --- containing only tuples
$(S,d)$ for which $S$ is finite. 

In this paper, we will focus only on aggregates with one set argument
but our theory can be extended easily to the more general case. An
example of an aggregate relation with two set arguments is the
generalized quantifier $\Aggr{most}\subseteq\PS{D}\times\PS{D}$ where
$\Aggr{most}(A,B)$ expresses that {\em most A's are B's}. The relation
$\Aggr{most}$ is defined as the set of all pairs of sets $(A,B)$ such
that $\Count(A\cap B)>\Count(A\setminus B)$. \label{PMost}

\subsection{Derived Aggregate Relations}

In this section we show how to obtain new aggregate relations by
composition of existing aggregates with other relations.

\begin{definition} 
  The {\em composition} of an aggregate relation
  $\R\subseteq\PS{D_1}\times D_2$ with a binary relation $P\subseteq
  D_2\times D_3$ is the aggregate relation
  $\R_P\subseteq\PS{D_1}\times D_3$ defined as: \[ \R_P = \{ (S,d)
  \vbar \exists d' \in D_2\colon (S,d')\in\R \text{ and }(d',d)\in
  P\}.\]
  
  The {\em composition} of an aggregate function $\F\colon \PS{D_1}\to D_2$
  with a binary relation $P\subseteq D_2\times D_3$ is the aggregate relation
  $\F_P\subseteq\PS{D_1}\times D_3$ defined as:
  \[ \F_P = \{ (S,d) \vbar (\F(S),d)\in P\}. \]
\end{definition}

Typically, the binary relation $P$ is some partial order relation on
the domain $D_2$. For example, the $\Count_\geq$ aggregate relation is
obtained as the composition of the $\Count$ aggregate function with
the $\geq$ relation on $\setN$ and contains all pairs $(S,n)$ such that
$\Count(S)\geq n$.

An aggregate relation can also be composed with a relation on
sets.  We consider only one instance of this sort of composition. 

\begin{definition} \label{def:aggr-subset}
  The {\em subset aggregate}  of an aggregate
  relation $\R \subseteq \PS{D_1}\times D_2$ is the aggregate
  $\R_\subseteq\subseteq \PS{D_1}\times D_2$ defined as:
  \[\R_\subseteq = \{(S,d)\vbar \exists S'\colon (S',d)\in\R \land S'\subseteq S\}. \]
\end{definition} 

For an aggregate function $\F$, $\F_\subseteq$ denotes the subset
aggregate of the graph of $\F$. As an example, $\Count_\subseteq(S,d)$
holds if for some subset $S' \subseteq S$, $\Count(S') = d$. In this
case, the two derived aggregates $\Count_\subseteq$ and $\Count_\geq$
are identical. This is because there exists a subset $S'\subseteq S$
such that $\Count(S') = d$ if and only if $\Count(S)\geq d$.

\subsection{Monotone and Anti-monotone Aggregates}

We define two different notions of monotonicity of aggregates, one for
functions and one for relations, and then show how they are related.

\begin{definition}
  Let $\langle D_2,\leq\rangle$ be a poset and $\F\colon \PS{D_1}\to
  D_2$ an aggregate function. We say that $\F$ is:
  \begin{itemize}
  \item {\em monotone} if $S_1\subseteq S_2$ implies $\F(S_1)\leq\F(S_2)$; 
  \item {\em anti-monotone} if $S_1\subseteq S_2$ implies 
        $\F(S_1)\geq\F(S_2)$.
  \end{itemize}
\end{definition}

The next two propositions list standard aggregate functions which are
monotone or anti-monotone with respect to some partial order. 

\begin{proposition} \label{prop:mon-aggr-func-inf}
  Let $\langle D,\leq\rangle$ be a complete lattice. The aggregate
  function $\Glb\colon \PS{D}\to D$ is anti-monotone with respect to
  $\leq$ and the aggregate function $\Lub\colon \PS{D}\to D$ is
  monotone with respect to $\leq$.
\end{proposition}
\begin{proof}
  Standard result from lattice theory \cite[Lemma 2.22]{Davey90}.
\end{proof}

In Table~\ref{tab:maf-fin} we use the following notation for subsets of
real numbers: $\setR^+$ for the set of non-negative numbers, $\setR^-$ for the
set of non-positive numbers, $\setR^{[1,\infty)}$ for the set of numbers in the
interval $[1,\infty)$, and $\setR^{[0,1)}$ for the set of numbers in the
interval $[0,1)$.

\begin{proposition} \label{prop:maf-fin}
  Let $D$ be a Cartesian product $D_1\times \dots \times D_n$ where $n\geq1$. The
  aggregate functions in Table~\ref{tab:maf-fin} from $\CFM(D)$, the
  set of finite subsets of $D$, to the poset $\struct{D',\leq}$ are
  monotone when $D_1$ is as given in the table.
\end{proposition}
\begin{proof}
  Follows immediately from well-known properties of real numbers.
\end{proof}

\begin{table}[htb]
  \[ \begin{array}{l|l|l}
    \text{aggregate} & D_1 & \struct{D',\leq}  \\
    \hline
    \Count &  arbitrary &  \struct{\setN,\leq} \\
    \Sum   & \setR^+     & \struct{\setR^+,\leq} \\
    \Sum   & \setR^-      & \struct{\setR^-,\geq} \\
    \Prod  & \setR^{[1,\infty)} & \struct{\setR^{[1,\infty)},\leq} \\
    \Prod  & \setR^{[0,1)}    & \struct{\setR^{[0,1)},\geq} \\
  \end{array} \]
  \caption{Monotone aggregate functions on finite sets}
  \label{tab:maf-fin}
\end{table}

Monotonicity and anti-monotonicity of an aggregate relation are defined
in the following way.

\begin{definition} \label{def:mon-aggr-rel}
  Let $\R\subseteq\PS{D_1}\times D_2$ be an aggregate relation. We say
  that $\R$ is:
  \begin{itemize}
  \item {\em monotone} if $(S_1,d)\in\R$ and $S_1\subseteq S_2$
    implies $(S_2,d)\in\R$;
  \item {\em anti-monotone} if $(S_2,d)\in\R$ and $S_1\subseteq S_2$
    implies $(S_1,d)\in\R$.
  \end{itemize}
\end{definition}

The next proposition summarizes the (anti-)monotonicity properties of
the aggregate relations defined in
Definition~\ref{Def-standard-aggr-relations}.

\begin{proposition} \label{prop:lb-ub}
  The aggregate relations $\Lb,\Ub\subseteq\PS{D}\times D$ on a poset
  $\langle D,\leq\rangle$ are anti-monotone.
\end{proposition}

All other relations defined in
Definition~\ref{Def-standard-aggr-relations} are neither a monotone nor
an anti-monotone.


We point out that the graph of a monotone aggregate function is not a
monotone aggregate relation according to
Definition~\ref{def:mon-aggr-rel}. For example, $\Lub$ is a monotone
aggregate function but its graph is not a monotone aggregate relation.
Instead, the composition of an aggregate function $\F$ with the
inverse of the order with respect to which it is monotone results in a
monotone aggregate relation.

\begin{proposition} \label{prop:aggr-comp-mono}
  Let $\F\colon\PS{D_1}\to D_2$ be an aggregate function which is monotone
  with respect to a partial order relation $\leq$ on $D_2$. Then $\F_\geq$
  and $\F_>$ are monotone aggregate relations and $\F_\leq$ and $\F_<$
  are anti-monotone aggregate relations.
\end{proposition}
\begin{proof}
  $\F_\geq$: Consider two sets $S_1$ and $S_2$ such that $S_1\subseteq S_2\subseteq D_1$
  and an element $d\in D_2$. Suppose that $\F_\geq(S_1,d)$ holds. By
  definition of $\F_\geq$ this is equivalent to $\F(S_1) \geq d$. Since $\F$
  is monotone we also have that $\F(S_1) \leq \F(S_2)$. So, we can
  conclude $\F(S_2)\geq d$ which is equivalent to $\F_\geq(S_2,d)$.

  The monotonicity of $\F_>$ and the anti-monotonicity of $\F_\leq$ and
  $\F_<$ can be proven in a similar fashion.
\end{proof}

As an application of this result we have that the aggregate relations
$\Count_\geq$ and $\Count_>$ are monotone since the aggregate function
$\Count$ is monotone with respect to $\leq$.

Forming the subset aggregate of any aggregate relation always results
in a monotone aggregate relation. 

\begin{proposition} \label{prop:aggr-subset-mono}
  Let $\R\subseteq \PS{D_1}\times D_2$ be an arbitrary aggregate relation. 
  \begin{enumerate}
  \item $\R_\subseteq$ is a monotone aggregate relation.
  \item If $\R$ is a monotone aggregate relation then $\R_\subseteq=\R$.
  \end{enumerate}
\end{proposition}

\section{First-order Logic with Aggregates}

We introduce aggregates in the context of many-sorted first-order logic.

A {\em sort symbol} (or simply sort) $s$ denotes some sub-domain of the
domain of discourse. A {\em product type} $s_1\times\dots\times s_n$ represents
the product of the domains represented by the sorts $s_1,\ldots,s_n$ and a
{\em set type} $\set{s}$ represents the set of all sets of elements
of sort $s$.

\begin{definition}[Aggregate Signature]
  An {\em aggregate signature} $\Sigma$ is a tuple 
        $\struct{ S;\; F;\; P;\; A}$  where
  \begin{itemize}
  \item $S$ is a set of {\em sorts};
  \item $F$ is a set of sorted {\em function symbols}  $f\colon s_1\times\cdots\times s_n\to w$
    where $n\geq0$;
  \item $P$ is a set of sorted {\em predicate symbols} $p\colon s_1\times\cdots\times s_n$
    where $n\geq0$; 
  \item $A$ is a set of sorted {\em aggregate symbols} $\R\colon
    \set{s_1\times\dots\times s_n}\times w$  where $n\geq1$.
  \end{itemize}
\end{definition}

We use $Sort(\Sigma)$, $Func(\Sigma)$, $Pred(\Sigma)$, and $Aggr(\Sigma)$ to denote the
sets $S$, $F$, $P$, and $A$ of $\Sigma$. We call a function symbol of the
form $f\colon \to w$ a {\em constant}. An aggregate symbol $\R\colon \set{s_1\times
  \dots \times s_n}\times w$ denotes an aggregate relation between sets of type
$s_1\times \dots \times s_n$ and objects of sort $w$. Of course, $Aggr(\Sigma)$ may
contain many instances of the same type of aggregate relation but with
different sorts.

For each sort $s$, we assume an infinite set $V_s$ of variables of
sort $s$ disjoint from the constants in $Func(\Sigma)$. We denote
variables, predicate symbols and function symbols with small letters
and constants with capital letters.

\begin{definition}[Terms and atoms]
  Let $\Sigma$ be an aggregate signature.  For every sort $s\in S$, we
  define the set of  {\em terms of type} $s$ by induction: 
\begin{itemize}
\item a variable $x\in V_s$ of sort $s$ is a term of type $s$;
\item if $f\colon s_1\times\dots\times s_n\to w\in Func(\Sigma)$ and
$t_1, \dots,t_n$ are terms of type $s_1, \dots, s_n$ respectively,
then $f(t_1,\dots,t_n)$ is a term of type $w$.
\end{itemize}
An {\em atom} has the form $p(t_1,\dots,t_n)$ where $p\colon
s_1\times\dots\times s_n\in Pred(\Sigma)$ is a predicate symbol and
$t_1,\dots,t_n$ are terms of types $s_1,\dots,s_n$ respectively.
\end{definition}

For a fixed aggregate signature $\Sigma$ we define the notions of set
expressions, aggregate atoms and formulas of the logic by simultaneous
induction.

\begin{definition}
  A {\em set expression} of type $\set{s_1\times\dots\times s_n}$ has the form
  $\se{(x_1,\dots,x_n)}{\varphi}$ where $\varphi$ is an aggregate formula called
  the {\em condition} of the expression and for each $i=1,\dots,n$,
  $x_i$ is a variable of sort $s_i$.
  
  An {\em aggregate atom} has the form $\R(s,t)$ where
  $\R\colon\set{s_1\times\dots\times s_n}\times w\in Aggr(\Sigma)$ is an aggregate symbol, $s$ is
  a set expression of type $\set{s_1\times\dots\times s_n}$, and $t$ is a term
  of type $w$.
  
  An {\em aggregate formula} is an atom, an aggregate atom, or an
  expression of the form $\lnot\varphi$, $\varphi\land \psi$, $\varphi\lor\psi$, $\forall x \varphi$ and $\exists
  x \varphi$ where $\varphi$ and $\psi$ are aggregate formulas and $x$ a
  variable. We also use $\varphi \subset \psi$ as an abbreviation for $\varphi \lor \lnot \psi$.
\end{definition}
The set of aggregate formulas over $\Sigma$ is denoted by $\CL^{aggr}_\Sigma$.

We illustrate the syntax of aggregate formulas with an example of
modeling power plant maintenance.

\begin{example}[Power Plant Maintenance] \label{ex:ppm}
  A power plant has a number of power generators called units 
  which have to be scheduled for  maintenance. 
  There is a restriction on the total capacity of the
  units in maintenance. Consider the following aggregate signature:
  \begin{multline*}
    \Sigma = \langle\{u,w,nat\};\; \{Max\colon nat\};\; \{capacity\colon u\times nat, maint\colon u\times w\times w\};\; \\ 
    \{\Sum\colon \set{nat\times u}\times nat \} \rangle.
  \end{multline*}
  The sort $u$ is interpreted with units and the sort $w$ with weeks.
  The predicate $capacity(u,c)$ represents  that a unit $u$ has a capacity
  $c$. The predicate $maint(u,s,e)$ specifies that unit $u$ is in
  maintenance during the period starting at time point $s$ (inclusive)
  and ending at time point $e$ (exclusive). The following aggregate
  formula expresses that the total capacity of the units in
  maintenance during a week, should not exceed a value $Max$:
  \[ \forall w\;
  \Sum_\leq(\se{(c,u)}{\exists s\exists e(maint(u,s,e)\land s\leq
    w<e\land capacity(u,c))}, Max). \] In this formula the sum
  aggregate computes the sum of all capacities $c$ of units $u$ that
  are in maintenance during week $w$. Note that each capacity $c$ is
  counted as many times as there are units $u$ with capacity $c$.
  \qed
\end{example}

A {\em positive literal} is an atom $p(t_1,\dots,t_n)$ and a {\em
  negative literal} is the negation of an atom $\lnot p(t_1,\dots,t_n)$.

An occurrence of a variable $x$ in an aggregate formula $\psi$ is {\em
  bounded} if it occurs in a subformula $\exists x \varphi$ or $\forall x \varphi$
of $\psi$ or in a set expression $\se{(x_1,\dots,x,\dots,x_n)}{\varphi}$ in
$\psi$. An occurrence of $x$ in $\psi$ is {\em free} if it is not bounded.
The set of free variables of $\psi$, denoted by $Free(\psi)$, is the set of all
variables with at least one free occurrence in $\psi$.
Terms and formulas without variables are called {\em ground} and those
without free variables are called {\em closed}.

Now, we define the semantics of the logic. Let $\Sigma$ be an
aggregate signature.

\begin{definition}[Structure]
  A {\em $\Sigma$-structure} $\CD$ consists of the following:
  \begin{itemize}
  \item for each sort $s\in S$ a domain $s^\CD$; 
  \item for each function symbol $f\colon s_1\times\dots\times s_n\to w\in Func(\Sigma)$ a
    function
    \[ f^{\CD}\colon s_1^\CD\times \dots \times s_n^\CD\to w^\CD; \]
  \item for each predicate symbol $p\colon s_1\times\dots\times s_n\in Pred(\Sigma)$ a
    relation
    \[ p^{\CD}\subseteq s_1^\CD\times\dots\times s_n^\CD. \]
  \item for each aggregate symbol $\R\colon\set{s_1\times\dots\times s_n}\times w\in Aggr(\Sigma)$ an
    aggregate relation $\R^{\CD}\subseteq \PS{s_1^\CD\times\dots \times s_n^\CD}\times w^\CD$.
  \end{itemize}
\end{definition}

Consider a $\Sigma$-formula $\varphi(x_1,\dots,x_n)$ with free variables
$x_1,\dots,x_n$ of sorts $s_1, \dots, s_n$, respectively and let
$d_1,\dots,d_n$ be elements of $s_1^\CD,\dots,s_n^\CD$, respectively.
Then, $\varphi(d_1,\dots,d_n)$ denotes the formula obtained by substituting
$d_i$ for each free occurrence of $x_i$ in $\varphi$. So, we consider domain
elements as new constants of the respective sorts. We denote this
enlarged signature with $\Sigma(\CD)$ and the corresponding set of formulas
with $\CL^{aggr}_{\Sigma(\CD)}$.


\begin{definition}
  The value $\eval{t}{\CD}$ of a ground term $t$ for a $\Sigma$-structure
  $\CD$ is defined inductively as follows:
  \begin{itemize}
  \item if $t$ is a domain element $d$, then $\eval{t}{\CD}=d$;
  \item if $t$ is a constant $c$, then $\eval{t}{\CD}=c^\CD$;
  \item if $t$ is a term $f(t_1,\ldots,t_n)$, then $\eval{t}{\CD} =
    f^{\CD}(\eval{t_1}{\CD},\ldots,\eval{t_n}{\CD})$. 
  \end{itemize}
\end{definition}


In the following definition and in the rest of the paper we often
treat a relation $R\subseteq D$ as a function $R\colon D \to \{\false,\true\}$ defined
as $R(d)=\true$ if and only if $d\in R$ for an element $d\in D$.

\begin{definition}[Truth function]\label{def:vf}
  Given is a $\Sigma$-structure $\CD$. We define the value
  $\eval{\se{\xx}{\varphi}}{\CD}$ of a set expression $\se{\xx}{\varphi}$ and the
  truth value $\CH_{\CD}(\psi)$ of an aggregate formula $\psi$ by
  simultaneous induction.

  The value $\eval{\se{(x_1,\dots,x_n)}{\varphi(x_1,\dots,x_n)}}{\CD}$ of a
  set expression is the set
  \[ \{(d_1,\dots,d_n)\in  s_1^\CD\times\dots\times s_n^\CD \vbar
    \CH_{\CD}(\varphi(d_1,\dots,d_n))= \true \} \] 
  
  The {\em truth function} $\CH_{\CD}(\cdot)\colon \CL^{aggr}_{\Sigma(\CD)}\to\C{TWO}$
  for closed aggregate formulas is defined in the following way:
  \[
    \begin{array}{lll}
      \CH_{\CD}(p(t_1,\dots,t_n)) &=  p^{\CD}(\eval{t_1}{\CD},\dots,\eval{t_n}{\CD}) \\
     \CH_{\CD}( \R(\se{\xx}{\varphi},t)) &=
  \R^\CD(\eval{\se{\xx}{\varphi}}{\CD},\eval{t}{\CD}) \\
      \CH_{\CD}(\lnot\varphi) &= \lnot\CH_{\CD}(\varphi) \\
      \CH_{\CD}(\varphi\lor\psi) &= \CH_{\CD}(\varphi) \lor \CH_{\CD}(\psi) \\
      \CH_{\CD}(\varphi\land\psi) &= \CH_{\CD}(\varphi) \land \CH_{\CD}(\psi) \\
      \CH_{\CD}(\exists x \varphi(x)) &= \bigvee_{d\in s^\CD}\CH_{\CD}(\varphi(d)) & \text{ (where $s$ the sort of $x$)}  \\
      \CH_{\CD}(\forall x \psi(x)) &= \bigwedge_{d\in s^\CD}\CH_{\CD}(\psi(d)) & \text{ (where $s$ the sort of $x$)}
    \end{array}
  \]
\end{definition}

We define $\CD\models\varphi$ if  $\CH_{\CD}(\varphi)=\true$. When $\CD\models\varphi$ we
call $\CD$ a {\em model} of $\varphi$. The relation $\models$ is called the truth
relation or the {\em satisfiability relation}. When the structure
$\CD$ is clear from the context, we drop the subscript $\CD$ from the
valuation function $\eval{\cdot}{}$ and truth function $\CH$.


We illustrate the use of first-order logic with aggregates to
formalize the well-known magic square problem.

\begin{example}[Magic Square] \label{ex:ms}
  Given is a $n\times n$ grid which has to be filled with the integer
  numbers from $1$ to $n^2$ such that the sum of the numbers in all
  rows, columns, and two diagonals is equal to the same number $M(n)$,
  known as the {\em magic constant}:
  \[ M(n) = \frac{n(n^2+1)}{2} \]
  Consider the following aggregate signature: 
  \begin{align*}
   \Sigma = \langle&\{pos, nat\};\; \{+\colon nat\times nat\to nat, *\colon nat\times nat\to nat, 
          /\colon nat\times nat\to nat,\\
    & Dim\colon nat, Mc\colon nat, f\colon pos\times pos\to nat\};\; \emptyset;\; \{\Sum\colon \{nat\}\times nat\} \rangle
  \end{align*}
  The sort $pos$ represents the positions of the table and the sort
  $nat$ the values of the table. The function symbol $f$ specifies the
  number in the corresponding row and column, the constant $Dim$ gives
  the dimension of the grid, and the constant $Mc$ gives the magic
  number. The problem is modeled by the following theory $T$:
  \begin{align*} 
    & Mc = Dim*(Dim*Dim+1)/2 \\
    & \forall x \forall y (1 \leq f(x,y)\leq Dim*Dim)\\
    & \forall x_1 \forall x_2 \forall y_1 \forall y_2 (f(x_1,y_1)=f(x_2,y_2) \to x_1= x_2 \land y_1=y_2)\\[10pt]
    & \forall y\; \Sum(\{z\vbar \exists x (z=f(x,y))\},Mc) \\
    & \forall x\; \Sum(\{z\vbar \exists y (z=f(x,y))\},Mc) \\
    & \Sum( \{ z \vbar \exists x (z=f(x,x)) \},Mc) \\
    & \Sum( \{ z \vbar \exists x (z=f(x,Dim+1-x)) \},Mc)
  \end{align*}
  Consider any structure $\CD$ such that $Dim^\CD = n \in \setN$,
  $pos^\CD=\{1,\dots,n\}$ and $nat^\CD = \setN$. Then $\CD$ is a model of
  $T$ if and only if $f^\CD$ specifies a solution for the magic square
  problem of dimension $n$. \qed
\end{example}

\section{Aggregate Programs} \label{sec:aggr-pr}

In this section, we define the syntax of aggregate programs and
introduce a basic semantical tool, the $T_P$ operator. 

Given are an aggregate signature $\Sigma$ 
and a set of sorted predicate symbols $\Pi$. We call the symbols from
$\Sigma$ {\em pre-defined} or {\em interpreted} while those from $\Pi$ {\em
  defined}. With $\Sigma(\Pi)$ we denote the aggregate signature consisting
of both sets of symbols.

From now until the end of this paper, we will assume a fixed aggregate
signature $\Sigma$ and a $\Sigma$-structure $\CD$ interpreting the pre-defined
symbols.

\begin{remark} 
Some of the pre-defined symbols are interpreted on standard domains like:
\begin{itemize}
\item sort symbols $nat$, $int$, $real$ interpreted by the sets of
  natural, integer and real numbers respectively;
\item the standard function symbols $+$, $*$, $-$, $\dots$
  on these sorts interpreted as the corresponding operations on numbers;
\item the standard predicate symbols $=$, $\leq$, $\dots$ on these sorts
  interpreted as the corresponding relations on numbers;
\item all aggregate symbols defined in Section~\ref{Sec-aggr}:
  $\Count$, $\Min$, $\Max$, $\Sum$, $\dots$.
\end{itemize}
Other interpreted symbols may be domain-specific.
In the context of logic programming, the interpretation of the set
$S_d\subseteq Sort(\Sigma)$ of domain-specific sorts and the set $F_d\subseteq Func(\Sigma)$ of
domain-specific function symbols is normally given by the {\em free
  term algebra} generated by $F_d$. The interpretations $s^\CD$ of all
sorts $s\in S_d$ and the interpretation $f^\CD$ of all function symbols
$f\colon s_1\times \dots\times s_n\to s \in F_d$ are defined by simultaneous induction as
follows:
\begin{itemize}
\item If $t_1\in s_1^\CD, \dots, t_n\in s_n^\CD$, then $f(t_1,\dots,t_n)\in s^\CD$. 
\item If $t_1\in s_1^\CD, \dots, t_n\in s_n^\CD$, then
  $f^\CD(t_1,\dots,t_n) = f(t_1,\dots,t_n)$.
\end{itemize} 
In case there is only one domain-specific sort, the free term algebra
corresponds to the Herbrand pre-interpretation, i.e., the Herbrand
universe and the Herbrand interpretation of function symbols.

The value of domain specific pre-defined predicate symbols may be
defined by an extensional database on the domain of $\CD$.  \qed
\end{remark}

A {\em $\Sigma(\Pi)$-aggregate rule} $r$ is of the form
\[ A \gets \varphi \] where $A$ is an atom of a defined predicate and
$\varphi$ is a $\Sigma(\Pi)$-aggregate formula. Note that $\varphi$
may contain universal quantifiers. The atom $A$ is called the {\em
  head} of the rule and the formula $\varphi$ the {\em body}.  We use
$body(r)$ to denote the body $\varphi$ of $r$. A {\em
  $\Sigma(\Pi)$-aggregate program} is a (possibly infinite) set of
aggregate rules.  A {\em normal aggregate program} is an aggregate
program in which the bodies of all rules are conjunctions of literals
and aggregate atoms.

Now, we introduce the basic semantical constructs. 

The {\em $\CD$-base} $base_{\CD}(\Pi)$ of $\Pi$ is defined as
\[
\begin{split}
  base_{\CD}(\Pi) = \{ p(d_1,\dots,d_n)\vbar & p\colon s_1\times\dots \times s_n\in\Pi, \text{ and } \\
                            & d_1\in s_1^\CD,\dots,d_n\in s_n^\CD\}. 
\end{split}
\]

The semantics of an aggregate program will be defined in the
collection of $\Sigma(\Pi)$-structures extending $\CD$.  For each subset $I$
of $base_{\CD}(\Pi)$, we define the $\Sigma(\Pi)$-structure $\CD(I)$ extending
$\CD$ such that for every atom $A\in base_{\CD}(\Pi)$: $\CH_{\CD(I)}(A) =
\true$ if and only if $A\in I$. Clearly, this is a one-to-one
correspondence between the subsets of $base_{\CD}(\Pi)$ and
$\Sigma(\Pi)$-extensions of $\CD$. In the rest of the paper, we exploit this
correspondence and use subsets of $base_{\CD}(\Pi)$, called {\em
  interpretations}, to represent $\Sigma(\Pi)$-extensions of $\CD$.
Sometimes, we also view an interpretation $I$ as a mapping $I\colon
base_{\CD} \to \C{TWO}$.

An interpretation $I$ is a {\em model} of an aggregate program $P$ if
$I$ is a model of the first-order theory obtained from $P$ by turning
every rule $A \gets \varphi$ into an implication $\forall \xx ( A \subset \varphi)$ where $\xx$
are the free variables of $A$ and $\varphi$.

The set $\CI=\PS{base_{\CD}(\Pi)}$ forms a complete lattice under the
subset order $\subseteq$.  This order extends to $\Sigma(\Pi)$-structures as follows:
$D(I)\leq D(J)$ if and only if $I\subseteq J$.

We introduce the following notation. For any closed defined atom
$A=p(t_1,\dots,t_n)$, $\eval{A}{\CD}$ denotes the atom
$p(\eval{t_1}{\CD},\dots,\eval{t_n}{\CD}) \in base_{\CD}(\Pi)$.

\begin{definition}
  The {\em instantiation} of a program $P$ over a structure $\CD$ is
  defined as the set $inst_{\CD}(P)$ of all closed rules $A \gets \varphi$
  such that:
  \begin{itemize}
  \item there exists a rule $A' \gets \varphi' \in P$ with free variables
    $x_1,\dots, x_m$ of sorts $s_1,\dots,s_m$, and
  \item there exist domain elements $d_1\in s_1^\CD,\dots, d_m\in
    s_m^\CD$, and
  \item $A=\eval{A'(d_1,\dots,d_m)}{\CD}$ and
  \item $\varphi = \varphi'(d_1,\dots,d_m)$.
  \end{itemize}
\end{definition}

Note that the body of a rule in the instantiation of an aggregate
program is a closed formula containing domain elements.

We now define the two-valued immediate consequence operator
of an aggregate program $P$.

\begin{definition} \label{def:tp-aggr}
  The {\em two-valued immediate consequence operator}
  $T_{P,\CD}^{aggr}\colon \CI\to\CI$ of an aggregate program $P$ is defined
  as: \[ T_{P,\CD}^{aggr}(I) = \{A \vbar A\gets \varphi \in inst_{\CD}(P) \text{
    and } \CD(I)\models \varphi\}. \]
\end{definition}
This operator extends the $T_P$ operator for normal logic programs
defined by \citeN{vanEmden76-JACM}. 

As for standard logic programs we have a correspondence between
models of an aggregate program $P$ and pre-fixpoints of
$T_{P,\CD}^{aggr}$.

\begin{proposition} \label{prop:pre-fp-mod}
  An interpretation $I$ is a model of an aggregate program $P$ if and
  only if $I$ is a pre-fixpoint of $T_{P,\CD}^{aggr}$, i.e.,
  $T_{P,\CD}^{aggr}(I) \leq I$.
\end{proposition}
\begin{proof}
  The proof is straightforward extension of the proof for standard
  logic programs.
\end{proof}

\begin{definition}
  An interpretation $I$ is a {\em supported model} of an aggregate
  program $P$ extending $\CD$ if $I$ is a fixpoint of
  $T_{P,\CD}^{aggr}$.
\end{definition}


Although the supported model semantics is generally considered to be a
weak semantics there are problems with aggregates for which it is the
appropriate semantics. One such example is the Party Invitation
problem \cite{Ross97-JCSS}.

\begin{example}[Party Invitation]\label{ex:pi-1}
  A number of people are invited to a party. A person $p$ will accept
  the invitation if and only if at least $k$ of his (her) friends also
  accept the invitation. Consider the following aggregate signature:
  \begin{align*}
    \Sigma = \langle&\{person, nat\};\; \{A\colon person, B\colon person\}; \\
     &\{thr\colon person\times nat, friend\colon person\times person \} ;\; \{\Count\colon \{person\}\times nat\} \rangle 
  \end{align*}
  and let 
  \[ \Pi= \{ accept\colon person\}. \] 
  Here $friend(x,y)$ means that $y$ is a friend of $x$ and $thr(x,t)$
  gives the lower bound $t$ on the number of friends of $x$. The
  problem can be modeled by the following single rule:
  \[ accept(x) \gets thr(x,t)\land\Count_\geq(\se{y}{friend(x,y)\land accept(y)},t).
  \] Consider an instance of the problem with two friends, say $A$
  and $B$. Each one of them accepts the invitation if and only if the
  other one accepts as well. This is represented by the $\Sigma$-structure
  $\CD$ in which $person^\CD=\{A,B\}$ and in which $friend^\CD$ and $thr^\CD$
  are given by the following table:
  \begin{align*} 
    &friend(A,B). & &thr(A,1). \\ 
    &friend(B,A). & &thr(B,1). 
  \end{align*} 
  The aggregate program has two supported models in which $accept$ is
  $\emptyset$ and $\{A,B\}$ respectively. The second solution is not minimal but
  it is a correct solution to the problem. In reality, $A$ and $B$ may
  communicate with each other about their decisions to attend the
  party. \qed
\end{example}

Several other examples for which the supported model semantics is
appropriate, including an elaborated version of the Party Invitation
problem, can be found in \cite{Pelov04-PhD}. It is worth noting that
each of these examples can also be expressed in first order logic with
aggregates using the completion of the aggregate program.



\subsection{Definite Aggregate Programs} \label{sec:aggr-def}

In the context of logic programming, definite logic programs are
negation free logic programs. A definite program $P$ characterizes a
monotone $T_P$ operator and its intended semantics is the least
fixpoint of $T_P$. In this section, we extend the notion of definite
program to programs with aggregates.

We define the notions of positive, negative, and neutral aggregate
formulas. This definition is not entirely syntactic, but also depends
on the monotonicity or anti-monotonicity of aggregate symbols
appearing in the formula. We can do that because aggregate symbols
always have a fixed interpretation given by the structure $\CD$.


\begin{definition}
  An occurrence of a predicate $P$ (resp. a formula $\psi$) in an
  aggregate formula $\varphi$ is {\em neutral}\/ if it occurs in the
  condition $\theta$ of an aggregate atom $\R(\{\xx\vbar \theta\}, t)$ in $\varphi$ such
  that $\R^{\CD}$ is neither monotone nor anti-monotone aggregate
  relation. Otherwise, the occurrence of $P$ (resp. $\psi$) is {\em
    positive} if the number of negations and aggregate atoms
  interpreted with an anti-monotone aggregate relation above $P$ (resp.\
  $\psi$) is even and {\em negative} if the number of negations and
  aggregate atoms interpreted with an anti-monotone aggregate
  relation above $P$ (resp.\ $\psi$) is odd.
\end{definition}


\begin{definition}[Positive and Negative Aggregate Formulas]\label{def:pn-f}
  An aggregate formula $\varphi$ is {\em positive} if no defined predicate
  occurs negatively or neutrally in $\varphi$. An aggregate formula $\varphi$ is
  {\em negative} if no defined predicate occurs positively or
  neutrally in $\varphi$.
\end{definition}

We note that in the above definition the polarity of pre-defined
symbols does not matter. Moreover, if a formula does not contain
defined atoms then it is both positive and negative. If the formula
$\varphi$ is an aggregate atom of the form $\R(\{\xx\vbar\varphi\},t)$ there are
three cases in which it can be positive. The first one is when
$\R^\CD$ is a monotone aggregate relation and $\varphi$ is a positive
formula. The second case is when $\R^\CD$ is an anti-monotone
aggregate relation and $\varphi$ is a negative formula. The third one is
when $\R^\CD$ is arbitrary and $\varphi$ does not contain defined
predicates. Similarly, the aggregate atom $\R(\{\xx\vbar\varphi\},t)$ is
negative if $\R^{\CD}$ is a monotone aggregate relation and $\varphi$ is
negative, $\R^{\CD}$ is an anti-monotone aggregate relation and $\varphi$ is
positive, or $\R^\CD$ is an arbitrary aggregate relation and $\varphi$ does
not contain defined symbols.

The main property of positive (resp.\ negative) aggregate formulas is
that their satisfiability is monotone (resp.\ anti-monotone) for a
given structure $\CD$.

\begin{proposition} \label{prop:mon-pos}
  Let $\CD$ be a $\Sigma$-structure and $\psi$ be a closed $\Sigma(\Pi)$-aggregate
  formula (possibly containing domain elements). For any pair $I \subseteq J \in
  base_{\CD}(\Pi)$, it holds that:
  \begin{itemize} 
  \item if $\psi$ is positive then $\CD(I)\models\psi$ implies $\CD(J)\models\psi$; 
  \item if $\psi$ is negative then $\CD(J)\models\psi$ implies $\CD(I)\models\psi$.  
  \end{itemize}
\end{proposition}
\begin{proof}
  The proof is by induction on the structure of $\psi$. For positive
  and negative formulas without aggregates this property is a standard
  result in the theory of first-order logic. We consider only the case
  when $\psi$ is an  aggregate atom
  $\R(\{\xx\vbar\varphi(\xx)\},t)$ without free variables. Let
  $S_I=\eval{\se{\xx}{\varphi(\xx)}}{\CD(I)}$ and
  $S_J=\eval{\se{\xx}{\varphi(\xx)}}{\CD(J)}$.
  
  First, let $\psi$ be a positive aggregate atom. We distinguish three
  cases.  
  \begin{enumerate} 
  \item $\R^{\CD}$ is a monotone aggregate relation and $\varphi(\xx)$ is a
    positive formula. For every well-sorted tuple $\dd$, $\varphi(\dd)$ is a
    positive formula as well. By the induction hypothesis, we have
    that $\CD(I)\models \varphi(\dd)$ implies $\CD(J)\models \varphi(\dd)$. Consequently, $S_I\subseteq S_J$.
    Finally, because $\R^{\CD}$ is a monotone aggregate relation,
    $(S_I,\eval{t}{\CD})\in\R^{\CD}$ implies $(S_J,\eval{t}{\CD})\in\R^{\CD}$.
    Thus, $\CD(I)\models \psi$ implies $\CD(J)\models \psi$.
  \item $\R^{\CD}$ is an anti-monotone aggregate relation and $\varphi(\xx)$
    is a negative formula. By the induction hypothesis, for every
    appropriate tuple $\dd$ of domain elements assigned to $\xx$, we
    have $\CD(J)\models \varphi(\dd)$ implies $\CD(I)\models\varphi(\dd)$.  Consequently,
    $S_J\subseteq S_I$. Finally, because $\R^{\CD}$ is an anti-monotone
    aggregate relation, $(S_I,\eval{t}{\CD})\in\R^{\CD}$ implies
    $(S_J,\eval{t}{\CD})\in\R^{\CD}$. Thus, $\CD(I)\models\psi$ implies $\CD(J)\models\psi$.
  \item If $\varphi$ contains no defined predicates, then $\CD(I)\models\psi$
    if and only if $\CD(J)\models\psi$ if and only if $\CD\models\psi$.
  \end{enumerate}
  
  The proof of anti-monotonicity of negative aggregate atoms is
  similar and is omitted.
\end{proof}

We point out that the class of positive (resp.\ negative) formulas are
a strict subset of the class for which the satisfiability relation is
monotone (resp.\ anti-monotone). For example the formula $p\lor\lnot p$ is a
tautology and hence it is monotone, however it is neither positive nor
negative.

\begin{definition} \label{def:dap}
  A {\em definite aggregate program} is an aggregate program such that
  the bodies of all rules are positive aggregate formulas.
\end{definition}

The class of definite aggregate programs is an extension of the class
of definite logic programs and has a monotone immediate consequence
operator.

\begin{theorem}\label{TDefiniteMono}
  If $P$ is a definite aggregate program then $T_{P,\CD}^{aggr}$ is monotone.
\end{theorem}
\begin{proof}
  Follows immediately from Proposition~\ref{prop:mon-pos}.
\end{proof}

\begin{definition}
  We define the {\em least fixpoint model} of a definite
  $\Sigma(\Pi)$-aggregate program $P$ extending $\CD$ as the least fixpoint
  of its immediate consequence operator $T_{P,\CD}^{aggr}$.
\end{definition}

A well-known example that can be modeled as a definite aggregate
program, is the Company Control problem
\cite{Kemp91-ILPS,Mumick90-VLDB,Ross97-JCSS,VanGelder92-PODS}.

\begin{example}[Company Control] \label{ex:cc}
  Given is a set of companies which own shares in each other. The
  problem is to decide if a company $x$ has a controlling interest in
  a company $y$. This is the case when $x$ owns (directly or through
  intermediate companies controlled by $x$) more than 50\% of the
  stock of $y$.
  
  To model the problem we use the following aggregate signature: \[ \Sigma
  = \langle\{c, s\};\; \emptyset;\; \{owns\_stock\colon c\times c\times s\};\; \{\Sum\colon \{s\times c\}\times s\}\rangle. \] The
  sort $c$ represents companies and the sort $s$ represents fractions
  of shares and is interpreted over the real interval $[0..1]$. The
  defined predicates are 
  \[ \Pi = \{ controls\colon c\times c\}. \]
  The predicate $owns\_stock(x,y,s)$ means that a company $x$ owns a
  fraction $s$ of the stock of a company $y$ and $controls(x,y)$ means
  that $x$ controls $y$. The problem is modeled by the aggregate
  program consisting of the following rule:
  \begin{align*}
    controls(x,y) \gets \Sum_>(\se{(s,z)}{&(x=z\lor controls(x,z))\land \\
      & owns\_stock(z,y,s)},0.5).
  \end{align*} 
  For numbers in the interval $[0..1]$, the $\Sum$ aggregate function
  is monotone with respect to $\geq$. Consequently, by
  Proposition~\ref{prop:aggr-comp-mono}, $\Sum_>$ is a monotone
  aggregate relation. Further, the formula in the aggregate atom is a
  positive formula, so the aggregate atom in the last rule is
  monotone.  Since none of the bodies contain negation this is a
  definite aggregate program with a monotone $T_{P,\CD}^{aggr}$
  operator which has a least fixpoint $I$.  \qed
%
\end{example}

We will now show that the least fixpoint of $T_{P,\CD}^{aggr}$
corresponds to the solution to the company control problem. We start
by giving a more precise definition of the control relation. Let
$sh(a,b)$ be a function which returns the fraction of shares of a
company $a$ in a company $b$ or $0$ if $a$ does not have shares in
$b$. We define for every $n\in\setN$ the {\em level $n$ control} binary
relation, denoted by $C^n$, by induction on $n$ as follows:
\begin{itemize}
\item $C^0=\emptyset$, i.e., no company has level 0 control of another company;
\item $C^{n+1}=\{(a,b)\vbar sh(a,b)+\sum_{(a,c)\in C^n} sh(c,b) > 0.5\}$ for
  $n\geq 0$, i.e., $a$ has a level $n+1$ control over company $b$ if the
  sum of the shares of $a$ in $b$ together with the shares of the
  companies which $a$ has level $n$ control in $b$ is more than 50\%.
\end{itemize}
Clearly, $C^n$ is an increasing sequence of relations. We define the
controls relation between companies $C$ as $C=\bigcup_{n\geq0} C^n$, i.e., a
company $a$ controls a company $b$ if, for some $n\geq0$, $(a,b)\in C^n$.

\begin{proposition}
  $controls^{\lfp(T_{P,\CD}^{aggr})}=C$.
\end{proposition}
\begin{proof}
  Let $I_n=T_{P,\CD}^{aggr}\uparrow^n(\emptyset)$ for $n\geq0$. We will prove for each
  $n\geq0$ that $controls^{I_n}= C^n$. Clearly, it follows from this that
  $controls^{\lfp(T_{P,\CD}^{aggr})}=C$.
  
  For $n=0$, $controls^{I_0}$ is empty and is equal to $C^n$. 
  
  For $n>0$, assume that $controls^{I_i}= C^i$ for $i=0,\ldots,n-1$. Fix
  two companies $a$ and $b$ and consider the value of the instance of
  the set expression:
  \[ S = \eval{\se{(s,z)}{(a=z\lor controls(a,z))\land owns\_stock(z,b,s)}}{I_{n-1}}. \]
  It is easy to see that if $controls^{I_{n-1}}=C^{n-1}$ then
  \[ S =\{ (s,c) \vbar (a,c)\in C^{n-1} \text{ and $c$ contains $s$
    shares in $b$}\}\cup S_1 \] where $S_1=\{(s,a)\}$ if $a$ has $s$ shares
  in $b$ and $S_1=\emptyset$ otherwise. It is straightforward then to see
  that $controls^{I_n}$ contains $(a,b)$ if and only if $(a,b)\in C^n$.
\end{proof}

\begin{example}[Borel Sets] \label{ex:bs}
  Let $\setR$ be the set of real numbers. {\em Borel sets} are defined by
  the following monotone inductive definition:
  \begin{itemize}
  \item any open set of real numbers is a Borel set;
  \item for any countable set $C$ of Borel sets, $\bigcap C$ and $\bigcup C$ are
    Borel sets;
  \item if $B$ is a Borel set then $\setR-B$ is a Borel set.
  \end{itemize}
  
  To model this definition as an aggregate program consider the
  following aggregate signature: 
  \[ \Sigma=\langle \{s\};\; \{compl\colon s\to s\};\; \{open\colon
  s\};\; \{\Glb^\omega_\subseteq, \Lub^\omega_\subseteq\colon
  \{s\}\times s\}\rangle. \] The $\Sigma$-structure $\CD$ interprets
  the sort $s$ with the set $\PS{\setR}$ of all subsets of the real
  numbers, the predicate $open$ is interpreted with the set of open
  sets, and the function $compl$ is interpreted as set complement:
  $compl^\CD(S)=\setR -S$.  The aggregate relations $\Glb^\omega$ and
  $\Lub^\omega$ are the restrictions of $\Glb$ and $\Lub$ to countable
  input sets, i.e., for any set $\C{R}$ of sets of real numbers and
  set $S$ of real numbers, $(\C{R},S)\in \Glb^\omega$ if and only if
  $\size{\C{R}}\leq \omega$ and $S=\bigcap \C{R}$. The aggregate
  relations $\Glb^\omega_\subseteq$ and $\Lub^\omega_\subseteq$ are
  obtained by forming the subset aggregates of $\Glb^\omega$ and
  $\Lub^\omega$ respectively (see Definition~\ref{def:aggr-subset}).
  Then $\Glb^\omega_\subseteq(\C{R},S)$ holds if $S$ is the
  intersection of some countable subset of $\C{R}$. Likewise
  $\Lub^\omega_\subseteq(\C{R},S)$ holds if $S$ is the union of some
  countable subset of $\C{R}$.
  
  The program defining Borel sets defines  a single defined
  predicate $borel\colon s$ and contains the following rules:
  \begin{align*}
    borel(S) & \gets open(S). \\
    borel(compl(S)) & \gets borel(S). \\
    borel(S) & \gets \Glb^\omega_\subseteq(\{B\vbar borel(B)\},S)\lor  \Lub^\omega_\subseteq(\{B\vbar borel(B)\},S). 
  \end{align*}
  
  Each of these rules is the formal representation of one of the rules
  in the inductive definition of Borel sets. Since $\Glb^\omega_\subseteq$ and
  $\Lub^\omega_\subseteq$ are monotone aggregate relations
  (Proposition~\ref{prop:aggr-subset-mono}) this is a definite
  aggregate program and it defines a monotone operator
  $T_{P,\CD}^{aggr}$. Consequently, the set of Borel sets is the least
  set of sets closed under the rules of the inductive definition and
  this corresponds exactly to the least fixpoint of
  $T_{P,\CD}^{aggr}$. Thus, $borel(d)\in \lfp(T_{P,\CD}^{aggr})$ if and
  only if $d$ is a Borel set. \qed
\end{example}

\subsection{Stratified Aggregate programs}\label{SStratified}

The important class of stratified aggregate programs was already considered
by several authors \cite{Mumick90-VLDB,DellArmi03-IJCAI,Faber04-JELIA}.
It is a natural extension of the concept of stratified logic program
\cite{Apt88-FDDLP} where aggregates are treated as negative literals.

\begin{definition}
An aggregate program $P$ is {\em stratified} if for each defined
predicate $p$, there is a unique natural number $\lm{p}>0$ called the
{\em level} of $p$ such that if $q$ occurs positively in the body $B$ of a rule
with head $p$, then $\lm{q}\leq \lm{p}$ and if $q$ occurs negatively in
$B$ or in an aggregate atom, then $\lm{q} < \lm{p}$. The level
$\lm{P}$ of $P$ is the maximum of the levels of the defined
predicates.
\end{definition}
 
For each level $i$, let $P_i$ be the set of all rules with a predicate
of level $i$ in the head and $\Pi_i$ the set of defined predicates
of level $i$. Define for each $i\geq 0$, $\Sigma_i= \Sigma \cup
\bigcup_{1\leq j \leq i}\Pi_j$.

Assume $i\geq 1$ and fix an arbitrary $\Sigma_{i-1}$-structure $\CD'$ extending $\CD$.
Notice that all predicates of $\Pi_i$ occur only positively in bodies of
$P_i$. Consequently, $P_i$ is a definite aggregate program and has a
monotone $T_{P_i,\CD'}^{aggr}$ operator. Note that it does not matter
whether the aggregates in $P_i$ are monotone or non-monotone, since they
do not contain predicates of $\Pi_i$.

\begin{definition}\label{DStrat}
  The {\em standard model} of an aggregate program $P$ extending $\CD$
  is the interpretation $I= \bigcup_{1\leq i \leq \lm{P}}I_i$ where the set $\{I_i
  \vbar 1\leq i \leq \lm{P}\}$ is defined by the following (finite)
  induction:
  \begin{align*}
    \CD_0 & = \CD;\\ 
    I_i & = \lfp(T^{aggr}_{P_i,\CD_{i-1}}); \\
    \CD_i & = \CD(\bigcup_{1\leq j \leq i}I_j).
  \end{align*}
\end{definition}

The aggregate program in the following example is a stratified
aggregate program.

\begin{example}[Shortest Path] \label{ex:sp-a}
  Consider the signature of directed weighted graphs
  \[ \Sigma=\langle \{n,w\};\; \emptyset;\; \{edge\colon n\times n\times w\};\; \{\Min\colon \{w\}\times w\} \rangle. \]
  A $\Sigma$-structure $\CD$ interprets the sort $n$ with a set of nodes,
  and the sort $w$, representing weights, with some set of real
  numbers $w^\CD\subseteq\setR$. The graph is defined by the relation $edge^\CD$
  where $(a,b,w)\in edge^\CD$ represents an edge from $a$ to $b$ with
  weight $w$.
  
  Consider the following formulation of the problem of finding the
  weight of the shortest path between two nodes which can be found in
  \cite[Example 4.1]{VanGelder92-PODS}.
  \begin{align*}
    & sp(x,y,w) \gets \Min(\se{c}{cp(x,y,c)},w). \\[10pt]
    & cp(x,y,c) \gets edge(x,y,c). \\
    & cp(x,y,c_1+c_2) \gets cp(x,z,c_1)\land edge(z,y,c_2).
  \end{align*}
  The aggregate relation $\Min$ is neither monotone nor anti-monotone,
  so the aggregate atom $\Min(\ldots)$ in the first rule is neutral.
  Consequently, the program is not definite. However, the program is
  stratified. The first stratum which defines the $cp/3$ predicate is
  a definite logic program. The predicate $cp/3$ represents the
  transitive closure of the graph: $cp(a,b,w)$ is true in the least
  model of $P_{cp}$ if and only if there is a path between $a$ and $b$
  with weight $w$. The second stratum contains only the definition of
  $sp/3$ and $sp(a,b,w)$ is true in the standard model of the program
  if and only if a shortest path between $a$ and $b$ exists and has
  weight $w$. \qed
\end{example}

\section{Ultimate Semantics for Aggregate Programs} \label{sec:ult-sem}

We start our study of the semantics of general aggregate programs with a 
brief investigation of  the semantics generated by the ultimate
approximating operator $U_{P,\CD}^{aggr}$ of $T_{P,\CD}^{aggr}$. This semantics of
aggregate programs was first studied by \citeN{Denecker01-ICLP}.


\begin{definition}
  The {\em ultimate approximating operator} $U_{P,\CD}^{aggr}\colon \CI^c\to\CI^c$ of
  $T_{P,\CD}^{aggr}\colon \CI\to\CI$ is defined as:
  \[ U_{P,\CD}^{aggr}(I_1,I_2) = 
      (\bigcap_{I\in [I_1,I_2]}\!\!\!T_{P,\CD}^{aggr}(I),
       \bigcup_{I\in [I_1,I_2]}\!\!\!T_{P,\CD}^{aggr}(I)\;). \]
\end{definition}

\begin{definition}
  The {\em ultimate Kripke-Kleene model}, the {\em ultimate
    well-founded model}, and the set of {\em ultimate stable models}
  of an aggregate program $P$ are defined as the Kripke-Kleene, the
  well-founded, and the set of exact stable fixpoints of the
  $U_{P,\CD}^{aggr}$ operator.
\end{definition}

\begin{example} \label{ex-ult-sem}
Consider the following program with Herbrand universe $\{0,1\}$:
\[ p(0)\gets \Count(\{x | p(x)\},1).\] Observe that this program has
two supported models: $\emptyset$ and $\{ p(0)\}$. Also, this is not a
definite aggregate program and  its immediate consequence
operator is non-monotone as can be seen from:
$$ T_{P,\CD}^{aggr}(\{p(0)\}) = \{p(0)\},$$
$$ T_{P,\CD}^{aggr}(\{p(0), p(1)\}) = \emptyset.$$
Let us construct the well-founded fixpoint. We start from the pair
$(\bot,\top)$.  The new upper and lower bounds are obtained by
applying the stable revision operator of Definition
\ref{def-stable-operator} on $(\bot,\top)=(\emptyset,\{p(0),p(1)\})$.
The new upper bound is the least fixpoint of
$(U_{P,\CD}^{aggr})^2(\emptyset,\cdot)$. It is easy to see that :
 $$
(U_{P,\CD}^{aggr})^2(\emptyset,\emptyset) = T_{P,\CD}^{aggr}(\emptyset) =\emptyset.
$$
It follows that $\emptyset$ is a fixpoint of this operator; it is
obviously the least fixpoint. Likewise, the new lower bound is the
least fixpoint of $(U_{P,\CD}^{aggr})^1(\cdot,\{p(0),p(1)\})$. This is
$\emptyset$ as well, since:
$$
(U_{P,\CD}^{aggr})^1(\emptyset,\{p(0),p(1)\}) = \emptyset;\\
$$
Consequently, the well-founded fixpoint is $(\emptyset,\emptyset)$.
This represents the two-valued interpretation $\emptyset$ and this is
also the unique ultimate stable model of this program.

The ultimate Kripke-Kleene model can be computed in two computation steps:
 $$\begin{array}{ll}
(U_{P,\CD}^{aggr})^1(\emptyset,\{p(0),p(1)\}) = \emptyset; & (U_{P,\CD}^{aggr})^2(\emptyset,\{p(0),p(1)\}) = \{ p(0)\} \\
(U_{P,\CD}^{aggr})^1(\emptyset,\{p(0)\}) = \emptyset; & (U_{P,\CD}^{aggr})^2(\emptyset,\{p(0)\}) = \{ p(0)\}.
\end{array}$$
The  model is the  three-valued interpretation $\{p(0)^\undef\} = (\emptyset,\{p(0)\})$. \qed
\end{example}

We obtain the following corollary to Theorem~\ref{TUltMono} and
Theorem~\ref{TDefiniteMono}.

\begin{corollary}\label{CUltMono}\label{CUltDefinite}
  If $T_{P,\CD}^{aggr}$ is monotone, then the ultimate well-founded fixpoint
  of $P$ is the least fixpoint of $T_{P,\CD}^{aggr}$ and the unique ultimate
  stable fixpoint of $P$. If $P$ is a definite aggregate program, then
  its ultimate well-founded model and unique ultimate stable model is
  the least fixpoint model of $P$.
\end{corollary}

It follows that both the ultimate well-founded and the ultimate stable
semantics correctly model the company control program in
Example~\ref{ex:cc} and the Borel sets program in Example~\ref{ex:bs}.
Later, we will also show a similar result to Corollary~\ref{CUltMono}
for stratified programs: if $P$ is a stratified program then its
ultimate well-founded and unique ultimate stable model coincide with
the standard model of $P$. Hence, the ultimate semantics also models
correctly the shortest path program in Example~\ref{ex:sp-a}.

Two aggregate programs with the same immediate consequence operator
are equivalent under ultimate semantics. Since substituting formulas
in rule bodies by equivalent formulas preserves the operator, this
operation is equivalence preserving.

\begin{proposition}\label{CTwoEquiv}
  Let $P$ and $P'$ be aggregate programs such that $P'$ is obtained by
  substituting a formula $\varphi'$ for a formula $\varphi$ in the body of a rule
  of $P$. If $\forall(\varphi\equiv\varphi')$ is satisfied in all two-valued
  $\Sigma(\Pi)$-extensions $\CD(I)$ of $\CD$, then $T_{P,\CD}^{aggr} =
  T_{P',\CD}^{aggr}$ and $P$ and $P'$ have the same ultimate
  Kripke-Kleene model, the same ultimate well-founded model, and the
  same set of ultimate stable models.
\end{proposition}

Another result about the set of ultimate stable models is that they
are always minimal models. In fact, we can prove such result for the
set of stable models associated with any approximating operator of
$T_{P,\CD}^{aggr}$.

\begin{proposition}
  Let $P$ be an aggregate program and $A$ be an approximating operator
  of $T_{P,\CD}^{aggr}$. Each stable model of $A$ is a minimal model
  of $P$.
\end{proposition}
\begin{proof}
  By Lemma~\ref{lem:mf-mm}, every stable model of $A$ is a minimal
  pre-fixpoint of $T_{P,\CD}^{aggr}$ and by
  Proposition~\ref{prop:pre-fp-mod}, the pre-fixpoints of
  $T_{P,\CD}^{aggr}$ are exactly the models of $P$.
\end{proof}

The nice semantical properties of ultimate semantics come at a
computational price. Even for programs without aggregates, computing
the ultimate well-founded model is \cc{co-NP}-hard and deciding the
existence of a two-valued ultimate stable model is
$\Sigma_2^p$-complete \cite{Denecker04-IC}.  For this reason, we will
study weaker semantics based on less precise approximations of
$T_{P,\CD}^{aggr}$. 

\section{Extending the Standard Well-founded and Stable Semantics}
\label{sec:std-sem}

The goal of this section is to extend the Kripke-Kleene
\cite{Fitting85-JLP}, well-founded \cite{VanGelder91-JACM} and stable
\cite{Gelfond88-ICSLP} semantics of normal logic programming.
According to Approximation Theory \cite{Denecker00-LBAI} these three
semantics can be obtained from the three-valued immediate consequence
operator $\Phi_P$ defined by \citeN{Fitting85-JLP}. In particular, the
collection of three-valued interpretations corresponds to the set
$L^c$ of consistent pairs of the lattice $L$ of two-valued
interpretations. The operator $\Phi_P$ is an approximation on this set,
it approximates the $T_P$ operator and its stable and well-founded
fixpoints correspond to the stable and well-founded models of $P$.  By
extending $\Phi_P$ to the class of aggregate programs, we will be able to
obtain well-founded and stable semantics which extend those of logic
programs without aggregates.

To extend the Fitting operator for aggregate programs, we must be able
to evaluate the aggregate formulas in three-valued interpretations.
For this reason we introduce the concept of a three-valued structure.
It is similar to standard structures, except that predicates are
assigned three-valued relations and aggregate symbols are assigned
three-valued aggregate relations. Because the value of a set
expression in a three-valued structure can be a three-valued set,
three-valued aggregates take three-valued sets as argument. We first
illustrate these points with an example.

\begin{example}
  We denote a three-valued set by indexing its {\em certain} elements
  with $\appr{\true}$ and its {\em possible} elements with
  $\appr{\undef}$. Let us fix the three-valued set $\appr{S}=
  \{1^{\appr{\true}},2^{\appr{\undef}},3^{\appr{\true}},5^{\appr{\undef}}\}$.

  A three-valued aggregate ${\C{C}\Aggr{ard}}$ of the cardinality
  aggregate $\Count$ is a mapping from pairs of
  three-valued sets and natural numbers to $\C{THREE}$. 
  In case of the set $\appr{S}$,  correct values for the set $\appr{S}$  are the following:
  \[ \left \{ \begin{array}{l}
      \C{C}\Aggr{ard}(\appr{S},n) = \appr{\undef}, \mbox{ for all $n\in \{2,3,4\}$}\\
      \C{C}\Aggr{ard}(\appr{S},n) = \appr{\false}, \mbox{ for all natural numbers $n\not
        \in \{2,3,4\}$}\end{array} \right . \] 
  This specifies that the set approximated by $\appr{S}$ has between two and
  four elements.

Similarly, correct values in case of $\Count_\geq$ are  as follows:
\[ \left \{ \begin{array}{l}
    \C{C}\Aggr{ard}_\geq(\appr{S},n) = \appr{\true}, \mbox{ for all $n\in \{0, 1, 2\}$}\\
  \C{C}\Aggr{ard}_\geq(\appr{S},n) = \appr{\undef}, \mbox{ for all $n\in \{3, 4 \}$}\\
\C{C}\Aggr{ard}(\appr{S},n) = \appr{\false}, \mbox{ for all natural numbers $n\not \in
\{0, \dots ,4\}$}\end{array} \right . \] This specifies that each set
approximated by $\appr{S}$ certainly has more than zero, one and two
elements, and has possibly more than three or four elements, but
definitely has not more than five  elements or more. A weaker but still correct value  for $\appr{S}$ would be:
\[ \left \{ \begin{array}{l}
    \C{C}\Aggr{ard}_\geq(\appr{S},0) = \appr{\true}, \mbox{ }\\
\C{C}\Aggr{ard}_\geq(\appr{S},n) = \appr{\undef}, \mbox{ for all $n\in \{1,2, 3, 4, 5, 6 \}$}\\
\C{C}\Aggr{ard}(\appr{S},n) = \appr{\false}, \mbox{ for all natural numbers $n>6$} \end{array} \right . \]
which specifies that the set approximated by $\appr{S}$ has certainly more than 0 elements, possibly more than one to six elements and certainly not more than 7 elements or more. \qed
\end{example}

We now formalize the notions of three-valued aggregate relations and
structures. 

\begin{definition}[Three-valued Aggregate Relations]
  A {\em three-valued aggregate relation} is a function $\CR\colon
  \PS{D_1}^c\times D_2\to \C{THREE}$ which satisfies:
  \begin{itemize}
  \item {\em $\leqp$-monotonicity}: for every pair of three-valued
    sets $\appr{S}_1, \appr{S}_2\in \PS{D_1}^c$ and for every $d\in D_2$,
    if $\appr{S}_1 \leqp \appr{S}_2$ then $\CR(\appr{S}_1,d) \leqp
    \CR(\appr{S}_2,d)$;
  \item {\em exactness}: for every exact (two-valued) set $S \in
    \PS{D_1}$ and for every $d\in D_2$, $\CR((S,S),d)\in \C{TWO}$. 
  \end{itemize}
\end{definition}

The concept of a three-valued aggregate relation is very similar to
approximating operators (Definition~\ref{def:ap}).

\begin{remark}
  The definition has a straightforward extension to aggregates with
  multiple set arguments by requiring $\leqp$-monotonicity and
  exactness conditions for all set arguments.
\end{remark}

A three-valued aggregate relation $\CR$ {\em approximates} an
aggregate relation $\R$ if for each set $S \in \PS{D_1}$ and for each
$d\in D_2$, $\CR((S,S),d) = \R(S,d)$. Due to the exactness condition, a
three-valued aggregate relation approximates exactly one aggregate
relation.

Recall that $\C{THREE}=\C{TWO}^c$. It follows that a three-valued
aggregate relation $\CR\colon \PS{D_1}^c\times D_2\to\C{THREE}$ is
completely determined by the pair $(\CR^1,\CR^2)$ of its projections
on the first and second component. These projections are relations
$\CR^1,\CR^2\subseteq \PS{D_1}^c\times D_2$ such that $\CR^1\subseteq
\CR^2$ and $\CR^1,\CR^2$ coincide on two-valued
sets\footnote{Equivalently, $\CR^1,\CR^2$ are functions
$\PS{D_1}^c\times D_2 \to \C{TWO}$ which coincide on two-valued sets
and such that $\CR^1 \leq \CR^2$.}. We will frequently define a
three-valued aggregate relation $\CR$ by defining $\CR^1$ and $\CR^2$
separately.

\begin{definition}[Three-valued Structure]
  Let $\Sigma$ be an aggregate signature. A {\em three-valued
    $\Sigma$-structure} $\ACD$ consists of the following:
  \begin{itemize}
  \item for each sort $s\in S$ a domain $s^{\ACD}$; 
  \item for each function symbol $f\colon s_1\times\dots\times s_n\to
    w\in Func(\Sigma)$ a function
    \[ f^{\ACD}\colon s_1^{\ACD}\times \dots \times s_n^{\ACD}\to w^{\ACD}; \]
  \item for each predicate symbol $p\colon s_1\times\dots\times s_n\in
    Pred(\Sigma)$ a {\em three-valued relation}
    \[ p^{\ACD}\colon  s_1^{\ACD}\times\dots\times s_n^{\ACD} \to \C{THREE}. \]
  \item for each aggregate symbol $\R\colon\set{s_1\times\dots\times s_n}\times w\in Aggr(\Sigma)$ a {\em
      three-valued aggregate relation} 
    \[ \R^{\ACD}\colon \PS{(s_1^{\ACD})^c\times\dots\times (s_n^{\ACD})^c}\times w^{\ACD} \to \C{THREE}. \]
  \end{itemize}
  
  A three-valued $\Sigma$-structure $\ACD$ {\em approximates} a
  $\Sigma$-structure $\CD$ if for each predicate symbol $p$, $p^{\ACD}$
  approximates $p^{\CD}$ and for each aggregate symbol $\R$,
  $\R^{\ACD}$ approximates $\R^{\CD}$.
\end{definition}

Now we define a precision order between three-valued aggregate
relations and structures.

\begin{definition}
  For all three-valued aggregate relations $\CR_1, \CR_2\colon \PS{D_1}^c\times
  D_2\to\C{THREE}$, define $\CR_1\leqp \CR_2$ if $\CR_1(\appr{S},d)\leqp
  \CR_2(\appr{S},d)$ for every three-valued set $\appr{S}\in\PS{D_1}^c$
  and domain element $d\in D_2$.
  
  For all three-valued $\Sigma$-structures $\ACD_1$ and $\ACD_2$, define
  $\ACD_1\leqp \ACD_2$ if $\ACD_1$ and $\ACD_2$ have the same domain,
  the same interpretations of sort and function symbols, for each
  predicate symbol $p\in Pred(\Sigma)$, $p^{\ACD_1} \leqp p^{\ACD_2}$ and for
  each aggregate symbol $\R\in Aggr(\Sigma)$, $\R^{\ACD_1} \leqp
  \R^{\ACD_2}$.
\end{definition}

It is straightforward to see that if $\CR_1\leqp \CR_2$ and $\CR_2$
approximates an aggregate relation $\R$ then $\CR_1$ also approximates
$\R$.


\begin{definition}[Three-valued valuation and truth functions]\label{def:pvf}
  Let $\Sigma$ be an aggregate signature and $\ACD$ be a three-valued
  $\Sigma$-structure. We define the {\em three-valued valuation function}
  $\eval{\cdot}{\ACD}$ for set expressions and the {\em three-valued truth
    function} $\CH_{\ACD}$ for aggregate formulas by simultaneous
  induction.
  
  Let $\se{(x_1,\dots,x_n)}{\varphi(x_1,\ldots,x_n)}$ be a set expression of
  type $\set{s_1\times\dots\times s_n}$.
  The value $\eval{\se{(x_1,\dots,x_n)}{\varphi}}{\ACD}$ is the three-valued
  set $\appr{S}$ defined as:
  \[ \appr{S}(d_1,\dots,d_n) = \CH_{\ACD}(\varphi(d_1,\ldots,d_n)) \]
  for every $(d_1,\dots,d_n)\in s_1^{\ACD}\times\dots\times s_n^{\ACD}$.

  The {\em three-valued truth function} for first-order aggregate
  formulas $\CH_{\ACD}(\cdot)\colon \CL^{aggr}_{\Sigma(\ACD)} \to \C{THREE}$ is
  defined as in Definition~\ref{def:vf} using the three-valued
  operations $\land$, $\lor$, and $\lnot$ as defined in Example~\ref{ex:three}.
%
\end{definition}

Next, we show that the three-valued truth function
$\CH_{\ACD}$ is monotone with respect to the precision order
$\leqp$ on three-valued interpretations.

\begin{proposition}\label{prop:ptf-mono} 
  Let $\ACD_1$ and $\ACD_2$ be three-valued $\Sigma$-structures. If $\ACD_1
  \leq_p \ACD_2$ then for every $\Sigma$-aggregate formula $\varphi$,
  $\CH_{\ACD_1}(\varphi)\leqp\CH_{\ACD_2}(\varphi)$.
\end{proposition}
\begin{proof}
  The proof is by a standard induction argument on the size of $\varphi$.
  For aggregate atoms it follows from the $\leqp$-monotonicity of the
  three-valued aggregate relations.
\end{proof}

Another proposition shows the correspondence between three-valued and
two-valued truth functions.  If all predicate symbols have two-valued
interpretations then evaluating an aggregate formula in a three-valued
structure results in a two-valued truth value.

\begin{proposition}\label{prop:ptf-extends} 
  Let $\CD$ be a $\Sigma$-structure and $\ACD$ be a three-valued structure
  which approximates $\CD$. For every aggregate formula $\varphi$ such that
  $p^{\ACD}$ is two-valued for all predicates $p$ appearing in $\varphi$,
  $\CH_{\ACD}(\varphi) = (\CH_{\CD}(\varphi),\CH_{\CD}(\varphi))$.
\end{proposition}
\begin{proof}
  The proof is by a standard induction argument on the size of $\varphi$.
  For aggregate atoms it follows from the exactness condition of the
  three-valued aggregate relations.
\end{proof}

In the sequel we will consider only three-valued structures for which
only the interpretation of the defined predicates $\Pi$ and the
aggregates is three-valued while the interpretation of the pre-defined
predicates is two-valued. Such structures are denoted by
$\ACD(\appr{I})$ where $\appr{I}\colon base_{\ACD}(\Pi)\to\C{THREE}$ gives the
(three-valued) interpretation of the predicates in $\Pi$.


We now extend Definition \ref{def:tp-aggr} of the immediate
consequence operator for aggregate programs to the three-valued case.

\begin{definition} \label{def:fp-aggr} The {\em three-valued immediate
    consequence operator} $\Phi_{P,\ACD}^{aggr}\colon \CI^c\to\CI^c$
  for an aggregate program $P$ maps any three-valued interpretation
  $\appr{I}$ to a three-valued  interpretation $\appr{I}' =
  \Phi_{P,\ACD}^{aggr}(\appr{I})$ such that for each ground defined
  atom $A\in base_{\ACD}(\Pi)$:
  \[ \appr{I}'(A) = \bigvee\{\CH_{\ACD(\appr{I})}(\varphi)\vbar
      A\gets \varphi\in inst_{\ACD}(P) \}. \] 
 \end{definition}
 Or, the truth value of a defined atom $A$ in $\appr{I}'$ is the
 greatest of all truth values of bodies of rule instances with $A$ in
 the head.
 


\begin{proposition}
  If $\ACD$ is a three-valued structure approximating $\CD$ then
  $\Phi_{P,\ACD}^{aggr}$ is an approximating operator of
  $T_{P,\CD}^{aggr}$.
\end{proposition}
\begin{proof}
  Follows from Proposition~\ref{prop:ptf-mono} and
  Proposition~\ref{prop:ptf-extends}.
\end{proof}

\begin{definition}
  Given a three-valued structure $\ACD$, the {\em $\ACD$-Kripke-Kleene
    model}, the {\em $\ACD$-well-founded model} and the set of {\em
    $\ACD$-stable models} of an aggregate program $P$ are defined as
  the Kripke-Kleene, well-founded and the set of exact stable
  fixpoints of the $\Phi_{P,\ACD}^{aggr}$ operator.
\end{definition}

\begin{example} 
  Reconsider the program of Example \ref{ex-ult-sem} with Herbrand
  universe $\{0,1\}$:
  \[ p(0)\gets \Count(\{x | p(x)\},1).\] We will show that, for an
  appropriate value of the three-valued aggregate, its standard
  well-founded model is identical to its ultimate well-founded model.

  To compute stable and well-founded models, we need to choose a
  three-valued aggregate $\C{C}\Aggr{ard}$ approximating $\Count$.
  Let us assume that
$$\C{C}\Aggr{ard}(\{0^{\appr{\undef}}, 1^{\appr{\undef}}\},1) = \C{C}\Aggr{ard}((\emptyset,\{0,1\}),1) = \appr{\undef} = (\false,\true)$$
$$\C{C}\Aggr{ard}(\{0^{\appr{\undef}}\},1) = \C{C}\Aggr{ard}((\emptyset,\{0\}),1) = \appr{\undef}= (\false,\true).$$
Then the following assignments can be computed easily:
 $$\begin{array}{lll}
(\Phi_{P,\ACD}^{aggr})^1 (\{p(0)^{\appr{\undef}}, p(1)^{\appr{\undef}}\}) & = (\Phi_{P,\ACD}^{aggr})^1((\emptyset,\{p(0), p(1)\}))  &= \emptyset;\\
(\Phi_{P,\ACD}^{aggr})^2 (\{p(0)^{\appr{\false}}, p(1)^{\appr{\false}}\}) & = (\Phi_{P,\ACD}^{aggr})^2((\emptyset,\emptyset))  &= \emptyset;\\
(\Phi_{P,\ACD}^{aggr})^2 (\{p(0)^{\appr{\undef}}, p(1)^{\appr{\undef}}\}) & = (\Phi_{P,\ACD}^{aggr})^2((\emptyset,\{p(0), p(1)\}))  &= \{p(0)\};\\
(\Phi_{P,\ACD}^{aggr})^1 (\{p(0)^{\appr{\undef}}\}) & = (\Phi_{P,\ACD}^{aggr})^1((\emptyset,\{p(0)\}))  &= \emptyset;\\
(\Phi_{P,\ACD}^{aggr})^2 (\{p(0)^{\appr{\undef}}\}) & = (\Phi_{P,\ACD}^{aggr})^2((\emptyset,\{p(0)\}))  &= \{p(0)\}.\end{array}$$
These assignments are the same as for the ultimate approximation $U^{aggr}_{P,\CD}$.  It follows that the empty set $\emptyset$ is the $\ACD$-well-founded model and  the unique $\ACD$-stable model of this program and  $\{p(0)^{\appr{\undef}}\}$ is the $\ACD$-Kripke-Kleene model. \qed
\end{example}

For logic programs without aggregates each $\Phi_{P,\ACD}^{aggr}$
operator coincides with the operator $\Phi_P$ defined by
\citeN{Fitting85-JLP}.
So, the well-founded and stable semantics of aggregate programs is an
extension of the well-founded and stable semantics of normal logic
programs.

Notice that the semantics of an aggregate program $P$ depends on
$\ACD$ and, in particular, on the choice of the three-valued
aggregates. This means that we still have a family of different
semantics. This family can be ordered by precision. Not surprisingly,
using more precise three-valued aggregates leads to more precise
semantics.

The following proposition is a straightforward consequence of
Proposition~\ref{prop:ptf-mono}.

\begin{proposition}\label{prop:ptf-aggr} 
  For every pair of three-valued $\Sigma$-structures $\ACD_1$ and $\ACD_2$
  and for every three-valued interpretation $\appr{I}$, if
  $\ACD_1\leqp\ACD_2$ then $\Phi_{P,\ACD_1}^{aggr}(\appr{I})\leqp
  \Phi_{P,\ACD_2}(\appr{I})$.
\end{proposition}


So, by Theorem~\ref{TPrec} we obtain the following result.

\begin{theorem}\label{prop:prec-sem}
  Let $P$ be an aggregate program and $\ACD_1$ and $\ACD_2$ be two
  three-valued $\Sigma$-structures such that $\ACD_1\leqp \ACD_2$. Then:
  \begin{itemize}
  \item the $\ACD_1$-Kripke-Kleene model of $P$ is less precise (in the
    $\leqp$ order) than the $\ACD_2$-Kripke-Kleene model of $P$;
   \item the $\ACD_1$-well-founded model of $P$ is less precise (in the
    $\leqp$ order) than the $\ACD_2$-well-founded model of $P$;
  \item every $\ACD_1$-stable model is a $\ACD_2$-stable model.
  \end{itemize}
\end{theorem}

The semantics that we have defined in this section do not satisfy all
the strong declarative properties of the ultimate semantics defined in
the previous section. For example, the $\ACD$-well-founded model of an
aggregate program with monotone immediate consequence operator is not
necessarily its least fixpoint. E.g. the program $\{ p \gets p\lor
\lnot p.\}$ has a constant, hence monotone $T_P$ with least fixpoint
$\{p\}$, but in its well-founded model $p$ is unknown. Also,
substituting a formula for an equivalent formula in a rule body is not
in general equivalence preserving. E.g. substituting $true$ for $p\lor
\lnot p$ in the above program does not preserve equivalence.  However,
some interesting properties  still hold. 

\begin{proposition}
  Let $P$ and $P'$ be aggregate programs such that $P'$ is obtained by
  substituting an aggregate formula $\varphi'$ for an aggregate formula $\varphi$
  in the body of a rule of $P$. If $\forall(\varphi\equiv\varphi')$ is satisfied in all
  three-valued $\Sigma(\Pi)$-extensions of $\ACD$, then $P$ and $P'$ are
  equivalent under the $\ACD$-Kripke-Kleene, $\ACD$-well-founded and
  $\ACD$-stable semantics.
\end{proposition}
\begin{proof}
  Follows from the fact that $P$ and $P'$ have the same three-valued
  immediate consequence operators.
\end{proof}

The three-valued equivalence condition in this proposition is strictly
stronger than the two-valued equivalence condition in
Proposition~\ref{CTwoEquiv}. For example $true$ and $p\lor \lnot p$ are
equivalent in two-valued semantics but not in three-valued.

 
Another important property is that in case of a stratified aggregate
program $P$, the $\ACD$-well-founded semantics and $\ACD$-stable
semantics coincide with the standard semantics as defined in
Section~\ref{SStratified}, and it does not matter how the aggregate
relations are approximated by $\ACD$.

\begin{theorem}\label{TStrat}
  Let $P$ be a stratified $\Sigma(\Pi)$-aggregate program. For any
  three-valued $\Sigma$-structure $\ACD$ approximating $\CD$, the
  $\ACD$-well-founded model is two-valued and is equal to the standard
  model of $P$ extending $\CD$ and to the unique $\ACD$-stable model
  of $P$.
\end{theorem}

The proof of this result depends on the following lemma.

\begin{lemma}\label{lem:rany-comp}
  Let $\varphi$ be a closed $\Sigma(\Pi)$-aggregate formula such that predicates of
  $\Pi$ do not occur in an aggregate atom.
  Let $\CD$ be a $\Sigma$-structure and $\ACD$ be a three-valued
  $\Sigma$-structure which is two-valued on all predicates in $\Sigma$ and
  approximates $\CD$. Then for any three-valued interpretation
  $(I_1,I_2)$, if the predicates in $\Pi$ occur only positively in $\varphi$
  then $\CH_{\ACD(I_1,I_2)}(\varphi)=(\CH_{\CD(I_1)}(\varphi),\CH_{\CD(I_2)}(\varphi))$
  and if the predicates in $\Pi$ occur only negatively in $\varphi$ then
  $\CH_{\ACD(I_1,I_2)}(\varphi)=(\CH_{\CD(I_2)}(\varphi),\CH_{\CD(I_1)}(\varphi))$.
\end{lemma}
\begin{proof}
  By simultaneous induction on the structure of $\varphi$. We give only the
  case when the predicates of $\Pi$ occur only positively in $\varphi$. The
  proof of the other case is symmetric.
  \begin{itemize}
  \item For a pre-defined atom $p(t_1,\ldots,t_n)$:
    \begin{align*}
      \CH_{\ACD(I_1,I_2)}(p(t_1,\dots,t_n)) &= \CH_{\ACD}(p(t_1,\dots,t_n)) \\
                 &= (\CH_{\CD}(p(t_1,\dots,t_n)),\CH_{\CD}(p(t_1,\dots,t_n))).
    \end{align*}
  \item For a user defined atom $p(t_1,\dots,t_n)$:
    \begin{align*}
      \CH_{\ACD(I_1,I_2)}(p(t_1,\dots,t_n)) 
         & = (I_1,I_2)(p(\eval{t_1}{\CD},\dots,\eval{t_n}{\CD})) \\
         & = (I_1(p(\eval{t_1}{\CD},\dots,\eval{t_n}{\CD})),
              I_2(p(\eval{t_1}{\CD},\dots,\eval{t_n}{\CD})))\\
         & = (\CH_{\CD(I_1)}(p(t_1,\dots,t_n)),
              \CH_{\CD(I_2)}(p(t_1,\dots,t_n))).
    \end{align*}
  \item For a formula with negation $\lnot\varphi$:
    \begin{align*}
       \CH_{\ACD(I_1,I_2)}(\lnot\varphi) & = \lnot\CH_{\ACD(I_1,I_2)}(\varphi) &
           & \text{by Definition~\ref{def:pvf}} \\
       & = \lnot(\CH_{\CD(I_2)}(\varphi),\CH_{\CD(I_1)}(\varphi)) 
       & & \text{by the induction hypothesis}\\
       & = (\lnot\CH_{\CD(I_1)}(\varphi),\lnot\CH_{\CD(I_2)}(\varphi))
       & & \text{by definition of $\lnot$} \\
       & = (\CH_{\CD(I_1)}(\lnot\varphi),\CH_{\CD(I_2)}(\lnot\varphi))
       & & \text{by Definition~\ref{def:vf}.}
    \end{align*}
  \item For a conjunction $\varphi\land\psi$:
    \[ \begin{split}
       \CH_{\ACD(I_1,I_2)}(\varphi\land\psi) & = 
       \CH_{\ACD(I_1,I_2)}(\varphi)\land\CH_{\ACD(I_1,I_2)}(\psi) \\
       & = (\CH_{\CD(I_1)}(\varphi),\CH_{\CD(I_2)}(\varphi)) \land
           (\CH_{\CD(I_1)}(\psi),\CH_{\CD(I_2)}(\psi)) \\
       & = (\CH_{\CD(I_1)}(\varphi)\land\CH_{\CD(I_1)}(\psi),
            \CH_{\CD(I_2)}(\varphi)\land\CH_{\CD(I_2)}(\psi)) \\
       & = (\CH_{\CD(I_1)}(\varphi\land\psi),\CH_{\CD(I_2)}(\varphi\land\psi)).
    \end{split} \]
  \item The proofs for formulas of the form $\varphi\lor\psi$, $\exists x \varphi$,
    and $\forall x \varphi$ are analogous.
  \item Let $\R(\{\xx\vbar\varphi\},t)$ be an aggregate atom. Since $\varphi$
    contains only pre-defined predicate symbols from $\Sigma$ then
    \[ \CH_{\ACD(I_1,I_2)}(\R(\{\xx\vbar\varphi\},t)) =
    \CH_{\ACD}(\R(\{\xx\vbar\varphi\},t)). \]
    Moreover, since the interpretation by $\ACD$ of all predicate symbols
    is two-valued we have (by Proposition~\ref{prop:ptf-extends}):
    \[ \CH_{\ACD}(\R(\{\xx\vbar\varphi\},t)) =
       (\CH_{\CD}(\R(\{\xx\vbar\varphi\},t)),\CH_{\CD}(\R(\{\xx\vbar\varphi\},t))). \]
  \end{itemize}
\end{proof}

\begin{proof}[Proof of Theorem \ref{TStrat} (Sketch)] 
Let $P$ be stratified by the level mapping $\lm{.}$, and let $\Pi_i$ be
the predicates of level $i$, $P_i$ the set of rules of $P$ with head in
$\Pi_i$.

  Given an interpretation $I\in \CI$, let us define $I|_i$ as the
  restriction of $I$ to the predicates of $\Pi_i$, and $I|_{\leq i}$ and
  $I|_{<i}$ as the restriction of $I$ to the predicates of $\bigcup_{j\leq
    i}\Pi_j$, respectively those of $\bigcup_{j<i}\Pi_j$. We extend these
  notations also to three-valued interpretations. It is easy to see
  that for every $i=1,\ldots,\lm{P}$ and every $\appr{I}, \appr{J} \in
  \CI^c$, if $\appr{I}|_{\leq i} = \appr{J}|_{\leq i}$, then
  \[ \Phi_{P,\ACD}^{aggr}(\appr{I})|_{\leq i} = \Phi_{P,\ACD}^{aggr}(\appr{J})|_{\leq i}. \] 
  Moreover, for every $\appr{I} \in \CI^c$:
  \[ \Phi_{P,\ACD}^{aggr}(\appr{I})|_i = \Phi_{P_i,\ACD(\appr{I}|_{<i})}^{aggr}(\appr{I}|_i). \]
  It follows from Corollary 3.12 in \cite{Vennekens-TOCL} that
  $\appr{I}$ is the well-founded fixpoint of $\Phi_{P,\ACD}^{aggr}$ if
  and only if for every $i=1,\ldots,\lm{P}$, $\appr{I}|_i$ is the
  well-founded fixpoint of $\Phi_{P_i,\ACD(\appr{I}|_{<i})}^{aggr}$.
  
  The next step is to prove by induction that for every
  $i=1,\ldots,\lm{P}$, the well-founded fixpoint $\appr{I}|_i$ of
  $\Phi_{P_i,\ACD(\appr{I}|_{<i})}^{aggr}$ is equal to $(I|_i,I|_i)$
  where $I$ is the standard model of $P$. This will show that the
  $\ACD$-well-founded model $\appr{I}$ of $P$ is equal to the standard
  model $I$ of $P$. Fix $i$ and assume that
  $\appr{I}|_{<i}=(I|_{<i},I|_{<i})$. Let $\Sigma_{i-1}=\Sigma\cup\bigcup_{j<i}\Pi_j$,
  $\CD'=\CD(I|_{<i})$ and $\ACD'=\ACD(\appr{I}|_{<i})$. Notice that
  $\ACD'$ approximates $\CD'$ and $\ACD'$ is two-valued on all
  predicates of $\Sigma_{i-1}$ (because $\appr{I}|_{<i}$ is a two-valued
  interpretation). Using Lemma~\ref{lem:rany-comp} we can show that
  for any three-valued $\Pi_i$-interpretation $(I_1,I_2)$,
  \[ \Phi_{P_i,\ACD'}^{aggr}(I_1,I_2) = (T_{P_i,\CD'}^{aggr}(I^1),T_{P_i,\CD'}^{aggr}(I^2)). \]
  By Theorem~\ref{TUltMono} it follows that $\Phi_{P_i,\ACD'}^{aggr}$ is
  also the ultimate approximation of $T_{P_i,\CD'}^{aggr}$. So, the
  $\ACD'$-well-founded model of $P_i$ is equal to the ultimate
  well-founded model of $P_i$ extending $\CD'$ and, by
  Corollary~\ref{CUltDefinite}, to the least fixpoint of
  $T_{P_i,\CD'}^{aggr}$.
\end{proof}

Since the $\Phi_{P,\ACD}^{aggr}$ operators are less precise than the
ultimate approximation $U_{P,\CD}^{aggr}$ we have the following
corollary of Theorems \ref{TPrec} and \ref{TStrat}.

\begin{corollary}
  For a stratified program $P$, the ultimate well-founded model is
  two-valued and is equal to the unique ultimate stable model and to
  the standard model of $P$.
\end{corollary}

In the next sections we define several different three-valued
aggregate relations and study the semantics obtained from the
corresponding $\Phi_{P,\ACD}^{aggr}$ operator. 

\subsection{Trivial Approximating Aggregates}

As a first example of a three-valued aggregate relation, we consider 
the least precise approximation of an aggregate. In the sequel, it will be
convenient to view an aggregate relation both as a subset of
$\PS{D_1}\times D_2$ and as a function from $\PS{D_1}\times D_2$ to $\C{TWO}$.

\begin{definition}[Trivial Approximating Aggregate]
  Let $\R\subseteq\PS{D_1}\times D_2$ be an aggregate relation. The
  {\em trivial approximating aggregate} $\triv{\R}\colon
  \PS{D_1}^c\times D_2\to\C{THREE}$ of $\R$ is defined as follows:
  \[ \triv{\R}((S_1,S_2),d) = 
  \begin{cases}
    (\false,\true) & \text{ if } S_1\neq S_2 \\
    (\R(S_1,d),\R(S_1,d)) & \text{ if } S_1=S_2
  \end{cases} \]
\end{definition}

\begin{proposition}
  For every aggregate relation $\R$, $\triv{\R}$ is a three-valued
  aggregate relation and $\triv{\R}$ approximates $\R$.
\end{proposition}

For a (two-valued) structure $\CD$ we define $\triv{\CD}$ as the
three-valued structure in which every aggregate relation $\R$ is
interpreted with $\triv{\R}$. When $\CD$ is clear from the context we
simply use $triv$.

As mentioned earlier, the trivial approximating aggregate is the least
precise three-valued aggregate relation.

\begin{proposition} 
  Let $\R$ be an aggregate relation. For every three-valued aggregate
  relation $\CR$ of $\R$, $\triv{\R}\leqp\CR$.
\end{proposition}

The trivial approximating aggregates are very imprecise. Nevertheless,
by Theorem~\ref{TStrat}, they suffice to model the semantics of the
important class of stratified aggregate programs.

\begin{corollary} 
  Let $P$ be a stratified aggregate program. The $triv$-well-founded
  model of $P$ is two-valued and is equal to the standard model of
  $P$ and the unique $triv$-stable model of $P$.
\end{corollary}

This corollary shows that even the weakest instance of the
well-founded and stable semantics suffices to define the standard
model approach for stratified aggregate programs.

\subsection{Ultimate Approximating Aggregate} \label{sec:aggr-ult}

In this section we investigate the instance of the $\Phi_P^{aggr}$
operator in which aggregate symbols are interpreted with the most
precise three-valued aggregate relation, called the {\em ultimate
  approximating aggregate}. This three-valued aggregate relation is
defined for all aggregate relations in a uniform way using a
construction similar to that of ultimate approximations. 

\begin{definition}[Ultimate Approximating Aggregate]
  Let $\R\subseteq\PS{D_1}\times D_2$ be an aggregate relation.  The {\em ultimate
    approximating aggregate} of $\R$ is a three-valued aggregate
  relation $\ult{\R}\colon \PS{D_1}^c\times D_2\to\C{THREE}$ defined as:
  \begin{align*}
    &\ult{\R}^1((S_1,S_2),d)=\true \text{ if and only if } 
    \forall S\in[S_1,S_2]\colon (S,d)\in \R \\
    &\ult{\R}^2((S_1,S_2),d)=\true \text{ if and only if }
    \exists S\in[S_1,S_2]\colon (S,d)\in \R
  \end{align*}
\end{definition}

\begin{proposition}
  For every aggregate relation $\R$, $\ult{\R}$ is a three-valued
  aggregate relation and $\ult{\R}$ approximates $\R$.
\end{proposition}

For a structure $\CD$ we define $\ult{\CD}$ as the three-valued
structure in which every aggregate relation $\R$ is interpreted with
$\ult{\R}$. When $\CD$ is clear from the context we simply use $ult$.

The ultimate approximating aggregate $\ult{\R}$ is the most precise in
the $\leq_p$-order among all possible three-valued aggregate relations.

\begin{proposition}\label{prop:uaa-prec} 
  Let $\R$ be an aggregate relation. For every three-valued aggregate
  relation $\CR$ which approximates $\R$, $\CR\leqp\ult{\R}$.
\end{proposition}

So, $\Phi_{P,ult}^{aggr}$ is the most precise operator in the
family of $\Phi_{P,\ACD}^{aggr}$ operators and by
Proposition~\ref{prop:prec-sem}, the $ult$-well-founded and the
$ult$-stable semantics are the most precise semantics of aggregate
programs in this family. Of course, these semantics are still weaker
than the ultimate well-founded and ultimate stable semantics from
Section~\ref{sec:ult-sem}.  Recall that for the program $\{ p \gets
p\lor \lnot p.\}$, $p$ is true in the ultimate well-founded model but
unknown in the $ult$-well-founded model. An aggregate program showing
similar behavior is given in the following example.
\begin{example} Consider the following program with Herbrand universe
  $\{0,1,2,3\}$:
  \[ p(x) \gets \Count_\leq(\{x | p(x) \},1) \lor \Count_\geq(\{x |
  p(x) \},2). \] Note that the body is a tautology in 2-valued
  logic. Therefore, in the ultimate semantics, this program is
  equivalent with:
  \[ p(x) \gets \true \] In the ultimate Kripke-Kleene, well-founded
  and unique stable model, $p$ is true for all domain elements. On the
  other hand, in the $ult$-well-founded model, each atom $p(i)$ has
  value $\appr{\undef}$ and the program has no $ult$-stable model. This can
  be seen as follows. First, we observe that for each three-valued
  interpretation $\appr{I}$, for every two-valued $I$ approximated by
  $\appr{I}$, $(\Phi_{P,ult}^{aggr})^2(\appr{I})$ is an upper bound
  of $T_{P,\CD}^{aggr}(I)=\{p(0),\dots,p(3)\} = \top$. Hence,
  $(\Phi_{P,ult}^{aggr})^2$ is a constant operator and maps each three-valued
  interpretation $\appr{I}$ to $\{p(0),\dots,p(3)\}$.  On the other
  hand, it is easy to see that for the three-valued set $\appr{S} =
  \{0^{\appr{\undef}},\dots,3^{\appr{\undef}}\}$, $\ult{\Card_\leq}^1(\appr{S},1)$ and
  $\ult{\Card_\geq}^1(\appr{S},2)$ are both false, since $\appr{S}$
  approximates sets with strictly more than one element and other sets
  with strictly less than two elements. Therefore,
  $$(\Phi_{P,ult}^{aggr})^1(\{p(0)^{\appr{\undef}},\dots,p(3)^{\appr{\undef}}\})= (\Phi_{P,ult}^{aggr})^1(\emptyset,\{p(0),\dots,p(3)\}) = \emptyset.$$
  Hence, $\emptyset$ is the least fixpoint of
  $(\Phi_{P,ult}^{aggr})^1(\cdot,\{p(0),\dots,p(3)\})$ and
  $\{p(0),\dots,p(3)\}$ is the least fixpoint of
  $(\Phi_{P,ult}^{aggr})^2(\emptyset,\cdot)$. It follows that the $ult$-well-founded fixpoint is 
  $$\{p(0)^{\appr{\undef}},\dots,p(3)^{\appr{\undef}}\} =
  (\emptyset,\{p(0),\dots,p(3)\}).$$ 

  As for the stable semantics, any $ult$-stable fixpoint is also a
  supported fixpoint, and the only supported fixpoint is
  $\{p(0),\dots,p(3)\}$. However, the least fixpoint of
  $(\Phi_{P,ult}^{aggr})^1(\cdot,\{p(0),\dots,p(3)\})$ is not
  $\{p(0),\dots,p(3)\}$ but $\emptyset$. Hence, there are no
  $ult$-stable fixpoints. \qed
\end{example}

Now we look at characterizations of the ultimate approximating
aggregate of some common aggregate functions.  Such characterizations
are useful for several purposes. First of all, they can be used in an
implementation of the semantics. Second, they can be used for
complexity analysis of the semantics.

For monotone and anti-monotone aggregates the truth value can be
computed directly on the boundary multisets.

\begin{proposition} \label{prop:uaa-mon}
  Let $\R\colon \PS{D_1}\times D_2$ be a monotone aggregate relation. Then
  $((S_1,S_2),d)\in\ult{\R}^1$ if and only if $(S_1,d)\in\R$ and
  $((S_1,S_2),d)\in\ult{\R}^2$ if and only if $(S_2,d)\in\R$.
\end{proposition}

\begin{proposition}\label{prop:uaa-anti}
  Let $\R\colon \PS{D_1}\times D_2$ be an anti-monotone aggregate
  relation. Then $((S_1,S_2),d)\in\ult{\R}^1$ if and only if
  $(S_2,d)\in\R$ and $((S_1,S_2),d)\in\ult{\R}^2$ if and only if
  $(S_1,d)\in\R$.
\end{proposition}

Next, we look at extrema aggregates, possibly defined on infinite
sets.

\begin{table}[htb]
  \[\begin{array}{l||l|l}
    \R & ((S_1,S_2),d)\in\ult{\R}^1 \text{ iff} & ((S_1,S_2),d)\in\ult{\R}^2 \text{ iff} \\
    \hline
    \Min   & d\in S_1 \land \Min(S_2,d)    & d\in S_2\land \lnot \exists d' \in S_1\colon d' < d\\
    \Max   & d\in S_1 \land \Max(S_2,d)    & d\in S_2\land \lnot \exists d' \in S_1\colon d' > d \\
    \Glb   & \Glb(S_1,d)\land\Glb(S_2,d) & \Glb(S_1\cup (S_2\cap[d,\top]),d) \\
    \Lub   & \Lub(S_1,d)\land\Lub(S_2,d)\; & \Lub(S_1\cup (S_2\cap[\bot,d]),d)
  \end{array} \]
  \caption{Characterization of ultimate approximating aggregates of
    extrema aggregates.} 
  \label{tab:alg-extr}
\end{table}

\begin{proposition}
  The characterizations in Table~\ref{tab:alg-extr} are correct.
\end{proposition}

Next, look at the aggregate functions $\Count$, $\Sum$ and $\Prod$
defined on finite sets.

\begin{proposition}[$\ult{\Count}$] \label{prop:alg-count}
  For every three-valued set $(S_1,S_2)$ and element $d$:
  \begin{align*}
    ((S_1,S_2),d)\in\ult{\Count}^1 & \text{ if and only if } 
    \size{S_1}=d=\size{S_2} \\
    ((S_1,S_2),d)\in\ult{\Count}^2 & \text{ if and only if } 
      \size{S_1} \leq d \leq \size{S_2}
  \end{align*}
\end{proposition}

\begin{proposition}[$\ult{\Sum}$ and $\ult{\Prod}$] \label{prop:sum-prod-1}
  For every three-valued set $(S_1,S_2)$ and element $d$:
  \begin{align*}
    ((S_1,S_2),d)\in\ult{\Sum}^1 \text{ iff } &
        \Sum(S_1,d)  
        \land \forall (x_1,\ldots,x_n)\in S_2\setminus S_1\colon x_1=0 \\
    ((S_1,S_2),d)\in\ult{\Sum}^2 \text{ iff } &  \exists S' \subseteq S_2\setminus S_1\colon \Sum(S_1\cup S',d) \\[12pt]
    ((S_1,S_2),d)\in\ult{\Prod}^1 \text{ iff } &
        \Prod(S_1,d) \land \\ 
          & ( d=0 \lor \forall (x_1,\ldots,x_n)\in S_2\setminus S_1\colon x_1=1) \\
    ((S_1,S_2),d)\in\ult{\Prod}^2 \text{ iff } & \exists S' \subseteq S_2\setminus S_1\colon \Prod(S_1\cup S',d).
  \end{align*}
\end{proposition}
\begin{proof}
  The proof is straightforward and uses only basic properties of
  numbers.
\end{proof}

Note that the definition of $\ult{\Sum}^2$ and $\ult{\Prod}^2$ is
simply a reformulation of the original definition of $ult$. In fact,
\citeN{Pelov04-PhD} shows that the complexity of computing
$\ult{\Sum}^2$ and $\ult{\Prod}^2$ is \cc{NP}-complete so it is
unlikely that any efficient algorithms can be found.

Finally, we look at combined aggregate relations of the form $\F_\geq$
and $\F_\leq$ where $\F\colon\CFM(D_1)\to D_2$ is an aggregate function on
finite sets and $\leq$ is a total order on $D_2$. For all three aggregate
functions $\Count$, $\Sum$, and $\Prod$ we can give efficient
algorithms for $\ult{\F_\geq}^1$ and $\ult{\F_\geq}^2$. We start with the
following general result. Let $\lmin_{\F},\lmax_{\F}\colon
\CFM(D_1)^c\to\CFM(D_1)$ be functions which for a given finite
three-valued set $(S_1,S_2)$ return a set $S\in[S_1,S_2]$ such that
$\F(S)$ is minimal (resp.\ maximal) over the set $[S_1,S_2]$, i.e.,
\begin{align*}
  &\forall S'\in[S_1,S_2]\colon \F(\lmin_{\F}(S_1,S_2)) \leq \F(S') \\
  &\forall S'\in[S_1,S_2]\colon \F(\lmax_{\F}(S_1,S_2)) \geq \F(S').
\end{align*}
Note that for a given aggregate function $\F$ and a three-valued
set $(S_1,S_2)$ there may be more than one set in the
interval $[S_1,S_2]$ with a minimal value of $\F$. For example
$\lmin_{\Sum}(\emptyset,\{0\})$ can return either $\emptyset$ or $\{0\}$.

The values of $\ult{\F_\leq}$ and $\ult{\F_\geq}$ can be computed using the
$\lmin_{\F}$ and $\lmax_{\F}$ functions in the following way.

\begin{proposition} \label{prop:caf-fm}
  Let $\F\colon \CFM(D_1)\to D_2$ be an aggregate function and $\leq$ be a total
  order on $D_2$. Then:
  \begin{align*}
    ((S_1,S_2),d)\in\ult{\F_\geq}^1 & \text{ if and only if } 
       \F(\lmin_{\F}(S_1,S_2))\geq d  \\
    ((S_1,S_2),d)\in\ult{\F_\geq}^2 & \text{ if and only if } 
       \F(\lmax_{\F}(S_1,S_2))\geq d
  \intertext{and similarly for $\ult{\F_\leq}$:}
    ((S_1,S_2),d)\in\ult{\F_\leq}^1 & \text{ if and only if } 
       \F(\lmax_{\F}(S_1,S_2))\leq d  \\
    ((S_1,S_2),d)\in\ult{\F_\leq}^2 & \text{ if and only if } 
       \F(\lmin_{\F}(S_1,S_2))\leq d.
  \end{align*}
\end{proposition}
\begin{proof}
  First note that, since $\{\F(S)\vbar S\in[S_1,S_2]\}$ is a finite totally
  ordered set, it always has a minimal and a maximal element. We give
  the proof for $\ult{\F_\geq}^1$:
  \begin{align*}
      & ((S_1,S_2),d)\in\ult{\F_\geq}^1  \\
    \Leftrightarrow & \forall S\in[S_1,S_2]\colon (S,d)\in\F_\geq  \\
    \Leftrightarrow & \forall S\in[S_1,S_2]\colon \F(S)\geq d   \\
    \Leftrightarrow & \forall x\in\{\F(S)\vbar S\in[S_1,S_2]\}\colon x\geq d \\
    \Leftrightarrow & \F(\lmin_{\F}(S_1,S_2))\geq d.
  \end{align*}
  The proofs of the other cases are analogous.
\end{proof}

So, to decide the first and second components of
$\ult{\F_\geq}((S_1,S_2),d)$ we need to compute the values
$\lmin_{\F}(S_1,S_2)$ and $\lmax_{\F}(S_1,S_2)$. First, we show how to
compute these two functions for any monotone or anti-monotone
aggregate function.

\begin{proposition} \label{prop:mm-mono}
  If $\F$ is a monotone aggregate function with respect to $\leq$ then
  $\lmin_{\F}(S_1,S_2) = S_1$ and $\lmax_{\F}(S_1,S_2) = S_2$. If $\F$
  is an anti-monotone aggregate function with respect to $\leq$ then
  $\lmin_{\F}(S_1,S_2) = S_2$ and $\lmax_{\F}(S_1,S_2) = S_1$.
\end{proposition}

This proposition can be applied to all aggregate functions listed in
Table~\ref{tab:maf-fin}.

For aggregate functions which are non-monotone the idea is to
partition the under and over-estimate of the input three-valued set to
subsets on which the aggregate function is monotone or anti-monotone.
Then we combine the sets on which the function is monotone to obtain
$\lmin_{\F}$ and the sets on which it is anti-monotone to obtain
$\lmax_{\F}$. We illustrate this idea for the $\Sum$ and $\Prod$
aggregate functions.

Below, $S^+$ denotes the set $\{(x_1,\ldots,x_n) \in S \vbar x_1 \geq 0\}$ and $S^-$ denotes
the set $\{(x_1,\ldots,x_n) \in S \vbar x_1 < 0\}$.
\begin{proposition}\label{prop:mm-sum}
  For every three-valued set $(S_1,S_2)$:
  \begin{align*}
    &\lmin_{\Sum}(S_1,S_2) = S_1^+ \cup S_2^-  \\
    &\lmax_{\Sum}(S_1,S_2) = S_1^- \cup S_2^+.
  \end{align*}
\end{proposition}
\begin{proof}
  Clearly the set $S'\in[S_1,S_2]$ with minimal sum is obtained by
  taking $S_1$ and all tuples with negative numbers from $S_2\setminus
  S_1$, that is
  \[ \lmin_{\Sum}(S_1,S_2)=S_1 \cup (S_2\setminus S_1)^-= S_1^+ \cup S_1^-\cup(S_2^-\setminus S_1^-) = S_1^+\cup S_2^-. \]
\end{proof}

The aggregate function $\Prod$ is non-monotone for sets
with arbitrary real numbers as first argument but is monotone for sets
with the first argument in the interval $[1,\infty)$ and anti-monotone
for sets with the first argument in the interval $[0,1]$. So, for
$\Prod$ on non-negative real numbers we can give a similar algorithm
as for $\Sum$ in Proposition~\ref{prop:mm-sum}.  Below,
$S^{[1,\infty)}$ denotes the set $\{(x_1,\ldots,x_n) \in S \vbar x_1
\in [1,\infty)\}$ and $S^{[0,1)}$ denotes the set $\{(x_1,\ldots,x_n)
\in S \vbar x_1 \in [0,1)\}$.


\begin{proposition} \label{prop:mm-prod-pos}
  For the aggregate function $\Prod^{\setR^+}\colon \CFM(\setR^+)\to\setR^+$,
  \begin{align*}
    &\lmin_{\Prod^{\setR^+}}(S_1,S_2) = S_1^{[1,\infty)} \cup S_2^{[0,1)} \\
    &\lmax_{\Prod^{\setR^+}}(S_1,S_2) = S_1^{[0,1)} \cup S_2^{[1,\infty)}.
  \end{align*}
\end{proposition}

The algorithms for $\Prod$ on the entire set of real numbers are more
complicated and can be found in \cite{Pelov04-PhD}.


As an application of the $ult$-well-founded semantics we consider a
recursive formulation of the shortest path problem.

\begin{example}[Shortest Path]\label{ex:sp-b} 
  Consider the following formulation of the problem of finding the
  shortest path \cite[Example 4.1]{VanGelder92-PODS}:
  \begin{align*}
    & sp(x,y,w) \gets \Min(\se{c}{cp(x,y,c)},w). \\[10pt]
    & cp(x,y,c) \gets edge(x,y,c). \\
    & cp(x,y,c_1+c_2) \gets sp(x,z,c_1)\land edge(z,y,c_2).
  \end{align*}
  The only difference between this program and the formulation of the
  shortest path in Example~\ref{ex:sp-a} is that we have replaced the
  $cp$ predicate in the body of the second clause of $cp$ with the
  $sp$ predicate. We have incorporated the knowledge that any
  shortest path of length $n+1$ must be an extension of a shortest
  path of length $n$. This fact is the basis of Dijkstra's algorithm.
  However, the program is no longer stratified because the $sp/3$
  predicate depends on itself through the $\Min$ aggregate relation
  which is non-monotone.
%
%
  \qed
\end{example}

It turns out the above program is only correct under certain
conditions on the graph. 

\begin{proposition}\label{prop:sp-b}
  Let $edge^\CD$ be a graph with the property that for any pair of
  nodes $a$ and $b$, if there is a path from $a$ to $b$, there is a
  shortest path from $a$ to $b$.  
  In the $ult$-well-founded model of the shortest path program from
  Example~\ref{ex:sp-b} an atom $sp(a,b,w)$ is:
  \begin{itemize}
  \item {\em true} if a shortest path between $a$ and
    $b$ exists and has weight  $w$;
\item {\em false} otherwise.
  \end{itemize}
\end{proposition}

The proof of the proposition and a deeper analysis of this program is
given in \ref{app}. There are several types of graphs for which we can
show that when there is a path from one node to another, there is a
shortest path between these nodes: acyclic finite graphs, finite
graphs with non-negative weights and infinite graphs with weights in
$\setN_0$. It follows from Proposition~\ref{prop:sp-b} that for these
types of graphs, the programs in Example~\ref{ex:sp-a} and
Example~\ref{ex:sp-b} are equivalent. 

There are also types of graphs which do not satisfy the
condition. Connected nodes without shortest path can occur if there is
a cycle with a negative weight between the two nodes. It can also
occur in infinite graphs.  An example of such a graph is $\{
(0,n+2,1), (n+2,1,1/n+2) \vbar n\in \setN \}$; although there are an
infinite number of paths from $0$ to $1$, there is no shortest path
between $0$ and $1$ (see \citeN{Ross97-JCSS} for another example). In
the appendix, we show that in such graphs, the well-founded model may
be three-valued or may even contain erroneous true $sp(a,b,w)$ atoms,
i.e., there are paths between $a$ and $b$ with strictly less weight
than $w$.

We conclude the section on ultimate approximating aggregates by
showing that for definite aggregate programs, the $ult$-well-founded
and $ult$-stable semantics are equal to the least fixpoint of the
$T_{P,\CD}^{aggr}$ which we defined in Section~\ref{sec:aggr-def}. The
key to the proof of this result is the following lemma.


\begin{lemma} \label{lem:tv-comp}
  Let $\ACD=ult(\CD)$ and let
  $(I_1,I_2)$ be a three-valued interpretation. If $\varphi$ is a closed
  positive aggregate formula then
  $\CH_{\ACD(I_1,I_2)}(\varphi)=(\CH_{\CD(I_1)}(\varphi),\CH_{\CD(I_2)}(\varphi))$. If
  $\varphi$ is a closed negative aggregate formula then
  $\CH_{\ACD(I_1,I_2)}(\varphi)=(\CH_{\CD(I_2)}(\varphi),\CH_{\CD(I_1)}(\varphi))$.
\end{lemma}
\begin{proof}
  The proof extends that of Lemma~\ref{lem:rany-comp} with several new
  cases when $\varphi$ is an aggregate atom $\R(\{\xx\vbar\psi\},t)$, $\psi$
  contains defined predicates and $\R^{\CD}$ is either a monotone or
  an anti-monotone aggregate relation. We consider only the case when
  $\R^{\CD}$ is a monotone aggregate relation and $\psi$ is a positive
  aggregate formula. The other three cases ($\R^{\CD}$ monotone and
  $\psi$ negative and $\R^{\CD}$ anti-monotone and $\psi$ positive or
  negative) are symmetric. First, note that since $\psi(\xx)$ is a
  positive formula then for every tuple of domain elements $\dd$,
  $\psi(\dd)$ is also positive. So,
  \[ \CH_{(I_1,I_2)}(\psi(\dd))= (\CH_{I_1}(\psi(\dd)),\CH_{I_2}(\psi(\dd))) \]
  and consequently
  \[ \eval{\{\xx\vbar\psi\}}{\ACD(I_1,I_2)}=
    (\eval{\{\xx\vbar\psi\}}{\CD(I_1)},\eval{\{\xx\vbar\psi\}}{\CD(I_2)}). \]
  We have:
  \begin{align*}
    \CH_{\ACD(I_1,I_2)}(\R(\{\xx\vbar\psi\},t)) 
    & = \ult{\R}(\eval{\{\xx\vbar\psi\}}{\ACD(I_1,I_2)},\eval{t}{}) \\
    & = \ult{\R}((\eval{\{\xx\vbar\psi\}}{\CD(I_1)},\eval{\{\xx\vbar\psi\}}{\CD(I_2)}),\eval{t}{}) \\
    \text{\small (by Proposition~\ref{prop:uaa-mon}) } 
    & = (\R(\eval{\{\xx\vbar\psi\}}{\CD(I_1)},\eval{t}{}),\R(\eval{\{\xx\vbar\psi\}}{\CD(I_2)},\eval{t}{})) \\
    & = (\CH_{\CD(I_1)}(\R(\{\xx\vbar\psi\},t)),\CH_{\CD(I_2)}(\R(\{\xx\vbar\psi\},t))).
  \end{align*}
\end{proof}

\begin{theorem}\label{thm:def-aggr-pr}
  Let $P$ be a definite aggregate program. Then $P$ has a two-valued
  $ult$-well-founded model $(M,M)$ which is also the single
  $ult$-stable model. Moreover $M=\lfp(T_{P,\CD}^{aggr})$.
\end{theorem}
\begin{proof}
  From Lemma~\ref{lem:tv-comp} follows that if $P$ is a positive
  aggregate program then
  \[ \Phi_{P,ult}^{aggr}(I_1,I_2)=(T_{P,\CD}^{aggr}(I_1),T_{P,\CD}^{aggr}(I_2)). \]
  By Theorem~\ref{TUltMono} follows that $\Phi_{P,ult}^{aggr}$ is
  also the ultimate approximation of $T_{P,\CD}^{aggr}$. So, the
  $ult$-well-founded model of $P$ is equal to the ultimate
  well-founded model of $P$ extending $\CD$ and, by
  Corollary~\ref{CUltDefinite}, to the least fixpoint of
  $T_{P,\CD}^{aggr}$.
\end{proof}

\subsection{Bound Approximating Aggregate}

The ultimate approximating aggregates have the disadvantage that they
may have a higher complexity than the aggregate relations which they
approximate. We already mentioned that the complexity of
$\ult{\Sum}^2$ and $\ult{\Prod}^2$ is \cc{NP}-complete
\cite{Pelov04-PhD} while the complexity of $\Sum$ and $\Prod$ is
polynomial. In this section we define a less precise approximating
operator for aggregate functions that are defined on totally ordered
finite sets of numbers.

\begin{definition}
  Let $\F\colon \CFM(D_1)\to D_2$ be an aggregate function and $\langle D_2,\leq\rangle$ be
  a totally ordered set. The {\em bound approximating aggregate}
  $\bnd{\F}\colon \CFM(D_1)^c\times D_2 \to \C{THREE}$ is defined as follows
  \begin{align*}
    ((S_1,S_2),d)\in \bnd{\F}^1 & \text{ if and only if } 
    \F(\lmin_\F(S_1,S_2)) = d = \F(\lmax_\F(S_1,S_2)) \\
    ((S_1,S_2),d)\in \bnd{\F}^2 & \text{ if and only if }
    \F(\lmin_\F(S_1,S_2)) \leq d \leq \F(\lmax_\F(S_1,S_2)).
  \end{align*}
\end{definition}

Note that by using Proposition~\ref{prop:caf-fm}, the definition of
$\bnd{\F}$ is equivalent to $\ult{\F_\geq}\land\ult{\F_\leq}$ where $\land$ is the
greatest lower bound in $\C{THREE}$ with respect to the $\leq$ order (see
Example~\ref{ex:three}). 

\begin{proposition}
  Let $\F\colon \CFM(D_1)\to D_2$ be an aggregate function and $\langle D_2,\leq\rangle$ be
  a totally ordered set. Then $\bnd{\F}$ is a three-valued aggregate
  relation of $\F$.
\end{proposition}

It is interesting to see how $\bnd{\F}$ compares to $\ult{\F}$. We
first show that the first components of the two three-valued aggregate
relations are always equal.

\begin{proposition}
  Let $\F\colon \CFM(D_1)\to D_2$ be an aggregate function and $\leq$ a total
  order on $D_2$. Then $\bnd{\F}^1=\ult{\F}^1$.
\end{proposition}

However, for most aggregate functions, $\bnd{\F}$ can be less
precise than $\ult{\F}$, i.e., $\bnd{\F}<_p\ult{\F}$ because $\bnd{\F}^2 >
\ult{\F}^2$. 

\begin{example}\label{ex:lps} 
  The pair $((\emptyset, \set{1,3}),2)$ is not in the relation $\ult{\Sum}^2$
  because there is no set $S\in[\emptyset,\set{1,3}]$ such that $\Sum(S)=2$. On
  the other hand $((\emptyset, \set{1,3}),2)\in\bnd{\Sum}^2$ because
  \begin{align*}
    b_1 = \Sum(\lmin_{\Sum}(\emptyset,\set{1,3})) & = \Sum(\emptyset) = 0, \\
    b_2 = \Sum(\lmax_{\Sum}(\emptyset,\set{1,3})) & = \Sum(\set{1,3}) = 4,
  \end{align*}
  and $b_1 \leq 2 \leq b_2$. So, $\ult{\Sum}^2\subset\bnd{\Sum}^2$ and
  consequently $\bnd{\Sum}<_p\ult{\Sum}$. \qed
\end{example}

\subsection{On the Complexity}

In this section we prove a simple complexity result. A full analysis
of the complexity of model generation for aggregate programs is beyond
the scope of this paper but we show that, despite the second order
nature of aggregates, model generation for aggregate programs may
remain tractable under an appropriate choice of the three valued
aggregates. 

The type of computational problem considered here is the model
extension problem \cite{Mitchell-Ternovska:AAAI2005}: given a signature $\Sigma(\Pi)$,
an aggregate program $P$ based on $\Sigma(\Pi)$ and an input $\Sigma$-structure
$\ACD$ which is two-valued on all predicates and three-valued on
aggregate symbols, compute the Kripke-Kleene model, the well-founded
model or an exact stable model.  Informally, we are interested here in
the complexity for instances of these problems with fixed $P$, fixed
$\Sigma(\Pi)$, and ``fixed'' interpretation of the aggregate symbols in
$Aggr(\Sigma)$ and for varying but finite input $\Sigma\setminus Aggr(\Sigma)$-structures
$\CD_o$. The problem with this intuition is that the interpretation
$\R^{\ACD}$ of a given aggregate symbol $\R$ in this class is of
course not really fixed: it varies with the input structure $\CD_o$. We
are interested in classes of problems where for example the sum
predicate is systematically interpreted by its ultimate approximating
aggregate, but evidently, the sum aggregate relation and its ultimate
approximating aggregate depend on the domain of the input structure
$\CD_o$.

\newcommand{\AgEx}{{X}}
\newcommand{\Cl}{{\C{C}}}

To circumvent this problem, we introduce the following concepts. Let
us fix an aggregate program $P$ based on signature $\Sigma(\Pi)$. Consider
the class $\Cl$ of two-valued structures of $\Sigma\setminus Aggr(\Sigma)$ with a finite
domain (i.e., with finite domains for every sort $s$). We assume a
given function $\AgEx$ from $\Cl$ to the class of $\Sigma$-structures such
that for each $\CD_o\in \Cl$, $\AgEx(\CD_o)$ extends $\CD_o$ with
three-valued aggregates $\CR$ for each aggregate symbol $\R\in Aggr(\Sigma)$.
Moreover, we assume that for each aggregate symbol $\R\in Aggr(\Sigma)$,
there are two Turing machines which, for (an appropriate encoding of)
arbitrary $\CD_o \in \Cl$ and arbitrary well-typed pair $(S,d)$
consisting of a three-valued set $\appr{S}$ and a domain element $d$,
compute respectively $(\R^{\AgEx(\CD_o)})^1(\appr{S},d)$ and
$(\R^{\AgEx(\CD_o)})^2(\appr{S},d)$.

Let $\AgEx(\Cl)$ be the image class of $\Cl$ under the function
$\AgEx$. The main result of this section is that if each three-valued
aggregate $\CR\in Aggr(\Sigma)$ is polynomially computable in the size of the
domain of the input structure, i.e., if for arbitrary $\CD_o$,
$\appr{S}$ and $d$, the Turing machines associated to $\CR$ can
compute $(\R^{\AgEx(\CD_o)})^1(\appr{S},d)$ and
$(\R^{\AgEx(\CD_o)})^2(\appr{S},d)$ in polynomial time in the size of
the domain of $\CD_o$ (i.e., the total number of elements in all
domains $s^{\CD_o}$ of all sorts $s$), then the following holds:

\begin{theorem}
\begin{itemize}
\item deciding if an atom $A$ is true in the Kripke-Kleene model of a
  program $P$ extending a structure in the class $\AgEx(\Cl)$ is in
  \cc{P};
\item deciding if an atom $A$ is true in the well-founded model of a
  program $P$ extending a structure in the class $\AgEx(\Cl)$ is in
  \cc{P};
\item deciding the existence of an exact stable model of a program
  $P$ extending a structure in the class $\AgEx(\Cl)$ is in \cc{NP}.
\end{itemize}
\end{theorem}

Note that, in the common case, computing the value of an aggregate
is polynomial in the size of the input three-valued set $\appr{S}$
(of a fixed type $s_1\times\dots\times s_n$). Then, since the number of
elements in such a set is bounded by a polynomial in the size of the
domain of $\CD_o$, computing the value of the aggregate is certainly
polynomial in the size of the domain of the input structure $\CD_o$.

\begin{proof}
  Let $L$ be a (finite) lattice and $A$ an approximation operator on $L^c$.
  Suppose that $n$ is the length of the longest chain in $L$. The
  computation of the Kripke-Kleene and well-founded fixpoint of $A$
  and the test whether a lattice element is an exact stable fixpoint
  of $A$ is done by monotone fixpoint computations.  It is easy to see
  that the number of applications of $A$ to compute its Kripke-Kleene
  fixpoint is bound by $n$.  Also testing whether a lattice element is
  an exact stable fixpoint of $A$ requires at most $n$ applications of
  $A$. Because the computation of the well-founded fixpoint involves
  an embedded fixpoint computation, its computation takes at most
  $n^2$ applications of $A$.
  
  Let us now consider a model extension problem for fixed $P,
  \Sigma(\Pi)$ and aggregate extension function $\AgEx$. Given an
  input $\Sigma\setminus Aggr(\Sigma)$-structure $\CD_o$, the lattice in which the
  computations take place is the power-set lattice
  $\CI=\PS{base_{\CD_o}(\Pi)}$. The maximal chain length in this lattice
  is the number of facts, i.e., the cardinality of $base_{\CD_o}(\Pi)$.
  This number is polynomial in the size of the domain of $\CD_o$. It
  follows then that all we need to prove to obtain the desired
  complexity results is that for any given pair $\appr{J} \in
  \C{\CI}$, we can compute $\Phi_{P,\AgEx(\CD_o)}^{aggr}(\appr{J})$ in
  polynomial time in the size of the domain.
  
  From Definition \ref{def:fp-aggr}, it follows that
  $\Phi_{P,\AgEx(\CD_o)}^{aggr}(\appr{J})$ corresponds to computing the
  truth value  of the bodies of all rule instances $A\gets \varphi\in
  inst(P)$.  It is clear that the number of rule instances is
  polynomial in the size of the domain of $\CD_o$. Therefore, all we need
  to prove is that for an arbitrary aggregate formula $\varphi[\xx]$
  with free variables $\xx$, the truth value of $\varphi[\xx/\dd]$ in
  $\appr{J}$, for arbitrary $\CD_o\in \Cl$ and tuple $\dd$ of domain
  elements, can be computed in polynomial time in the domain size of
  $\CD_o$.
  
  In case $\varphi$ is a first order formula, the polynomial
  computability of its truth value with respect to a three-valued
  structure is proven by induction on the structure of $\varphi$.  We
  need to extend this proof with the additional case that $\varphi$ is
  an aggregate atom. Computing the truth value of an aggregate atom
  $\R(s,d)$ requires firstly, to evaluate the contained set expression
  $s$ and compute its three-valued set $\appr{S}$, and secondly, to evaluate
  the truth value of $(\CR^{\AgEx(\CD_o)})^1(\appr{S},d)$ or
  $(\CR^{\AgEx(\CD_o)})^2(\appr{S},d)$. It follows straightforwardly from the
  induction hypothesis that the first task can be done in polynomial
  time in the domain size of $\CD_o$, whereas the second task is
  polynomial by assumption.
\end{proof} 

Our main motivation for developing the semantic framework of this
section was the high complexity of the ultimate well-founded and
stable semantics as defined in Section \ref{sec:ult-sem}. This result
shows that under appropriate choice of the three-valued aggregates, we
indeed obtain weaker but tractable Kripke-Kleene and well-founded
semantics. 

The above theorem is subject to a strong limitation, in particular the
restriction to {\em finite} structures. Many of the interesting
applications of aggregates (including the company control and the
shortest path problem) contain integer or real numbered domains.
Frequently used aggregates such as $\Sum$ and $\Count$ range over
these infinite domains. Clearly, in the context of infinite domains,
only strong syntactical conditions on the form of rules can guarantee
termination or tractability of the model generation process. But this
is the case whether the program contains aggregate expressions or not.
To discover such conditions is an important topic for future
research but it is beyond the scope of this paper to investigate this
issue. 

\subsection{Summary of the results}

In this and the previous section, we have introduced a family of
Kripke-Kleene, a family of stable and a family of well-founded
semantics for aggregate programs, parameterized by the approximation
operator. We introduced also a sub-family of these semantics, obtained
by extending the standard three-valued Fitting operator with different
three-valued aggregates.  We presented two generic ways for deriving a
three-valued aggregate from a given (two-valued) aggregate relation,
called the {\em trivial} and the {\em ultimate} three-valued
aggregate.

All instances in each of the three families are {\em consistent} with
each other. In particular, when one instance of the Kripke-Kleene or
well-founded semantics infers that a literal is true, there is no
instance in which this literal is false and any more precise instance
of the same semantics will infer the same literal. Also, a model in
one instance of the stable semantics is also a model in each more
precise instance of the stable semantics. This shows that all
instances of each of the three families of semantics basically
formalize the same intuitions but with different degree of precision,
with the semantics based on the ultimate operator as the most precise.
However, for several important subclasses of aggregate programs,
optimal precision is reached using weaker approximations. As a general
rule, when the well-founded model of one approximation is two-valued,
then it coincides with the well-founded and unique stable model of
each more precise approximation. In Theorem \ref{thm:def-aggr-pr} and
{Corollary} \ref{CUltMono}, we proved that for definite aggregate
programs, the least fixpoint semantics, the $ult$-well-founded,
$ult$-stable, the ultimate well-founded and the ultimate stable
semantics all coincide. One example is the Company Control Example
\ref{ex:cc}. For stratified aggregate programs, Theorem \ref{TStrat}
showed that the standard model, the $triv$-well-founded and
$triv$-stable model and the ultimate well-founded and stable model all
coincide. This class was illustrated by the first Shortest Path
Example \ref{ex:sp-b}. In case of the second Shortest Path Example
\ref{ex:sp-b}, Proposition \ref{prop:sp-b} showed that (under
some weak conditions) the $ult$-well-founded model is two-valued.

This raises the question of what a good choice of the
semantics is and how we can exploit our results to build an effective
reasoning system for aggregate programs.  Just as for standard logic
programming, the family of semantics of aggregate programs offers a
trade-off between {\em precision} versus {\em complexity}.  Therefore,
choosing the ``right'' semantics is a {\em pragmatic} matter and the
relevant question is what degree of precision is required for the
applications that one has in mind.

In the case of logic programs without aggregates, the standard and the
ultimate semantics only differ for programs with a rare combination of
reasoning by cases and recursion, as in $\{ p \gets p \lor \lnot p \}$.
We are not aware of a single non-artificial example of a logic
program, appearing in the literature, which shows this behaviour.
This means that, de facto, the standard and the ultimate semantics
coincide and the ultimate semantics can be computed using the more
efficient Fitting operator. In case of aggregate programs, the
situation seems even better. It must be admitted that applications of
recursion over aggregates in the literature are quite rare.  We
believe that this is not a coincidence. Indeed, aggregates provide a
rather powerful way to avoid recursion.  In a language without
aggregates, the means to compute the value of some aggregate of a set
of objects (e.g., cardinality, sum, minimum or maximum, \dots) is
often by using recursion over the potential elements of the set
\cite{VanGelder92-PODS}. By allowing aggregates in the language, such
applications of recursion can be avoided. Recall that for stratified
aggregates programs even the $triv$-stable and $triv$-well-founded
semantics are sufficiently precise. As for the applications of
recursion over aggregates in this paper, the Party Invitation program
of Example \ref{ex:pi-1} is to be interpreted under the supported
model semantics, whereas for the second Shortest Path Example
\ref{ex:sp-b} and the Company Control Example \ref{ex:cc}, the
$ult$-well-founded model is the correct solution.

In summary, we believe that extensions of the standard semantics with
polynomially computable three-valued aggregates, provide sufficient
precision for almost all practical applications. This is the case for
the ultimate approximating aggregates of $\Card$, $\Min$, $\Max$ and
the bound approximating aggregates of $\Sum$ and $\Prod$
\cite{Pelov04-PhD}. To us, it seems that the most suitable
approximation for being incorporated in model generation or query
systems for aggregate programs, is the extension of the Fitting
operator with those three-valued aggregates.

\section{Related Work} \label{sec:rw}

Aggregate relations are closely related to generalized quantifiers
\cite{Lindstrom66}. An example of a generalized quantifier is
$\Aggr{most}(A,B)$, defined on page \pageref{PMost}, which expresses that
most elements of set $A$ belong to set $B$. Clearly, this relation can
be viewed as a binary aggregate relation with two set arguments.
Standardly, the notion of generalized quantifier is formalized in a
slightly more involved way than as a second order predicate in an
arbitrary domain. Instead the concept is formalized as a class of
structures closed under isomorphism. For example, $\Aggr{most}$ could
be formalized as the class of all structures consisting of a domain
$D$ and a binary relation $M \subseteq \PS{D}\times \PS{D}$ consisting of all pairs
$(A,B)$ such that $A$ is finite and more than half of the elements of
$A$ belong to $B$. In this way, a domain independent characterization
of the generalized quantifier is obtained. Aggregates could be
formalized similarly. For example, the cardinality aggregate could be
formalized as the class of all structures consisting of a domain $D$
and a binary relation $C \subset \PS{D}\times \setN$ containing all tuples
$(S,n)$ such that $S$ is a subset of $D$ containing $n$ elements.
An extensive study of generalized quantifiers in three-valued logic is
done by \citeN{vanEijck96}. The notion of a {\em super-valuation
interpretation} of a generalized quantifier, as defined there,
coincides with our notion of ultimate approximating aggregate of the
corresponding aggregate relation.


In the context of logic programming, many different approaches to
aggregates have been proposed. Below, we discuss a selection of
them. 

The class of {\em  monotonic aggregate programs} \cite{Mumick90-VLDB} is very
similar to the class of definite aggregate programs. A monotonic
program is a program in which every rule is monotonic. A monotonic
rule is a rule $r$ such that $I\models body(r)$ and $I\subseteq J$
implies $J\models body(r)$ for any pair of interpretations $I$ and
$J$. Although this is a semantic definition of monotonicity, the
authors introduce a sufficient syntactic condition for monotonicity of
a rule. Essentially, an aggregate atom can appear only in formulas of
the form
\begin{equation}
  \label{eq:mon-lit}
  \exists z \R(\se{\xx}{q(\xx)},z)\land p(z,t)
\end{equation}
where $p$ is a pre-defined relation. Moreover, the satisfiability of
this formula must be monotone. In our syntax \eqref{eq:mon-lit} can be
expressed as the aggregate atom 
\begin{equation}
  \label{eq:aa-dar}
  \R_P(\se{\xx}{q(\xx)},t)
\end{equation}
using the derived aggregate relation $\R_P$. Then, the condition that
the satisfiability of \eqref{eq:mon-lit} is monotone is equivalent to
the condition that $\R_P$ is a monotone aggregate relation (and
consequently \eqref{eq:aa-dar} is a positive aggregate atom). The
notion of positive aggregate atoms is simpler and, in our opinion,
more natural than the condition of monotonic literals of
\cite{Mumick90-VLDB}.

A common approach to extend the stable semantics of general logic
programs with negation to aggregate programs was to treat aggregate
literals as negative literals when computing the program reduct
\cite{Kemp91-ILPS,Gelfond02-CL,Elkabani04-ICLP}. Such semantics is
relatively easy to define and the definition also extends to answer
sets of disjunctive logic programs. However, it has been shown that
this semantics accepts non-minimal models and does not model correctly
some problems with recursion over aggregation, for example the Company
Controls program from Example~\ref{ex:cc}
\cite[Section~4.3.3]{Pelov04-PhD}. For aggregate programs containing
only negative aggregate literals it has been shown
\cite{Pelov04-LPNMR} that the set of $ult$-stables models is the same
as the set of stable models defined by the above authors. 


A more elaborate definition of a stable semantics was given by
\cite{Simons02-AI} for programs with weight constraints and implemented
by the well-known smodels system. In our language, weight constraints
correspond to aggregate atoms build with the $\Sum_\leq$ and $\Sum_\geq$
aggregate relations. An extensive comparison between the $ult$-stable
semantics and the stable semantics of weight constraints can be found
in \cite{Pelov04-LPNMR,Pelov04-PhD} and will not be repeated here.

A novel feature of the language of weight constraints was that it
allows weight constraints to be present also in the head of the rules.
This approach have been further developed in different directions.
One line of research was to consider different variations and
extensions of weight constraints like abstract constraints
\cite{Marek04-AAAI}, monotone cardinality atoms
\cite{Marek04-LPNMR:mca} or set constraints \cite{Marek04-LPNMR:sc}.
Such constraint atoms correspond in a natural way to aggregate atoms.
The stable semantics of these extensions is also defined in terms of
lattice operators. However, since constraint atoms are allowed in the
heads of rules, the operators become non-deterministic and the
algebraic theory is quite different than the approximation theory we
used in this work. However, all the semantics agree on the class of
definite aggregate programs and its least model semantics. The
equivalent of a definite logic program in \cite{Marek04-LPNMR:sc} is
called a Horn SC-logic programs and such programs are also
characterized by a unique model which is the least fixpoint of a
deterministic monotone operator $S_P$ which is the equivalent of our
$T_{P,\CD}^{aggr}$ operator.

Another extension of the language of weight constraint atoms is to
considers arbitrary propositional formulas containing arbitrary
aggregate atoms (both in the head and in the body of the rule)
\cite{Ferraris05-LPNMR}. The answer set semantics for such
propositional formulas is different than the $\ACD$-stable semantics
which we defined. The main reason is that the semantics of
\cite{Ferraris05-LPNMR} is based on the logic of here-and-there
\cite{Lifschitz01-TOCL} while the $\ACD$-stable semantics are based on
Kleene's strong three-valued logic $\C{THREE}$. The simplest example
of the difference between the two logics is that $\lnot \lnot p$ is equivalent
to $p$ in $\C{THREE}$ while it is not in the logic of here-and-there.
As a consequence, the program consisting of the single rule $p \gets \lnot\lnot p$
is a definite program according to Definition~\ref{def:dap} because $\lnot
\lnot p$ is a positive (aggregate) formula. So, all our semantics assign
to this program the model $\emptyset$ On the other hand, it has two models $\emptyset$
and $\{p\}$ according to the answer set semantics of
\cite{Lifschitz01-TOCL,Ferraris05-LPNMR}. This difference also
manifests for aggregate formulas. For example, it is easy to see that
the aggregate formulas $\lnot \Sum_{\neq}(s,t)$ and $\Sum(s,t)$ have the same
three-valued truth value for any interpretation of $\Sum$ with a
three-valued aggregate relation while they are not equivalent under
the semantics of \cite{Ferraris05-LPNMR}.


Another proposal for a stable semantics of disjunctive logic programs
extended with aggregates was given in \cite{Faber04-JELIA}. In the
sequel we investigate in more detail the relationship with this
semantics to the family of $\ACD$-stable semantics defined earlier.
First, we recall the definitions of the stable semantics of
\cite{Faber04-JELIA}.

\begin{definition}[\citeNP{Faber04-JELIA}]\label{def:dlp-st}
  The {\em reduct} $P^I$ of an aggregate program $P$ with respect to
  an interpretation $I$ is a program obtained from $inst(P)$ by
  deleting all rules in which a literal or an aggregate atom in the
  body is false in $I$.
  An interpretation $I$ is a {\em FLP-stable model} of $P$ if $I$ is a
  minimal model of $P^I$.
\end{definition}

In one direction we can show the following result.

\begin{proposition}\label{prop:flp}
  For any aggregate program $P$ and for any three-valued structure
  $\ACD$, if an interpretation $I$ is an $\ACD$-stable model of $P$
  then $I$ is a FLP-stable model of $P$.
\end{proposition}
\begin{proof}
  We will show that any $\ACD$-stable model $I$ is also a
  $\ACD$-stable model of $P^I$.  It will then follow from
  Lemma~\ref{lem:mf-mm}, that $I$ is a minimal pre-fixpoint of
  $T_{P^I,\CD}^{aggr}$.  Since the pre-fixpoints of
  $T_{P^I,\CD}^{aggr}$ are exactly the models of $P^I$, this will
  imply that $I$ is a minimal model of $P^I$ and hence, $I$ is a
  FLP-stable model of $P$. We start by showing
  \begin{equation}
    \label{eq:fpi}
      \forall J\subseteq I\colon (\Phi_{P^I,\ACD}^{aggr})^1(J,I)=(\Phi_{P,\ACD}^{aggr})^1(J,I). \tag{*}
  \end{equation}
  This will imply that
  $\lfp((\Phi_{P^I,\ACD}^{aggr})^1(\cdot,I))=\lfp((\Phi_{P,\ACD}^{aggr})^1(\cdot,I))=I$,
  so $I$ will be a stable model of $P^I$. To show \eqref{eq:fpi} we
  only need to look at the rules $r\in P-P^I$ and show that
  $\CH_{\ACD(J,I)}^1(body(r))=\false$. The definition of reduct
  implies that for such a rule $\CH_{\CD(I)}(body(r))=\false$ where
  $\CD$ is the (two-valued) structure approximated by $\ACD$. We also
  have that $(J,I)\leq_p(I,I)$ which implies that
  \[ \CH_{\ACD(J,I)}^1(body(r)) \leq \CH_{\ACD(I,I)}^1(body(r)) = \CH_{\CD(I)}(body(r)) =
  \false \]
  So, $\CH_{\ACD(J,I)}^1(body(r))=\false$.
\end{proof}

The next example gives a program for which the two semantics disagree.

\begin{example} \label{ex:dlp}
  Consider the following aggregate program $P$:
  \begin{align*}
    r & \gets \Count_{\neq}(\se{x}{p(x)},1). \\
    p(A) & \gets r. \\
    p(B) & \gets r. \\
    p(A) & \gets p(B). \\
    p(B) & \gets p(A).
  \end{align*}
  The program has only one model $M=\{r,p(A),p(B)\}$.
  This is also a FLP-stable model because $P^M=P$ and $M$ is also a
  minimal model of $P$. However, the program does not have an ultimate
  total stable model and, consequently, it does not have a total stable
  model for any less precise approximating operator. 

To illuminate what is going on in this example, let us make the
following observation. In the context of the Herbrand universe
$\{A,B\}$, the aggregate atom $\Count_{\neq}(\se{x}{p(x)},1)$
expresses that $p$ has either zero or two elements. Or, in each
two-valued Herbrand interpretation this atom is equivalent to
\begin{equation}\label{eq:AB}
  (\lnot p(A) \land \lnot p(B)) \lor (p(A) \land p(B)).
\end{equation}
In fact, if we interpret $\Count_{\neq}$ by the ultimate approximating
aggregate, then a simple case analysis shows that, in each
three-valued Herbrand interpretation, the truth values of the
aggregate atom and of the formula \eqref{eq:AB} are
identical\footnote{The three-valued equivalence of $\ult{\Count_{\neq}}$
  and \eqref{eq:AB} is an application of a general transformation of
  ultimate approximations of aggregate atoms to formulas \cite[Section
  5.3.6]{Pelov04-PhD}.}.  It follows that the program $P$ and the
program obtained by substituting the rules
\begin{align*}
  r & \gets \lnot p(A) \land \lnot p(B). \\
  r & \gets  p(A) \land p(B). 
\end{align*}
for the first rule in $P$, have identical three-valued immediate
consequence operators and hence, have identical stable
models\footnote{Stronger, both three-valued and two-valued immediate
  consequence operators coincide, and hence, all semantics of the two
  programs based on these operators coincide, including
  $ult$-Kripke-Kleene, $ult$-stable and $ult$-well-founded
  semantics.}. The second program is a standard logic program and it
is easy to see that it has no stable models. In particular, if we
compute the reduct under $M=\{r,p(A),p(B)\}$, only the second new rule
remains and together with the rest of the rules the least model is
$\emptyset$. Thus $M$ is not a stable model of the translated program.  \qed
\end{example}
  
The above example illustrates a natural principle of the semantics
defined in our framework. Substituting an aggregate atom by an
aggregate free formula which is equivalent with respect to two-valued
semantics, preserves ultimate well-founded and ultimate stable models.
Substituting an aggregate atom by an aggregate free formula which is
equivalent with respect to three-valued semantics preserves the
standard well-founded and stable semantics. As shown by the example,
this natural principle is not satisfied by the semantics defined in
\cite{Faber04-JELIA}. This is a fundamental weakness of this semantics. 

%
%
%
%

Finally, we mention another recent work on defining a stable semantics
of aggregates \cite{Son07-TPLP}. As shown in that paper, the semantics
defined there coincides with the $ult$-stable semantics and provides
an interesting different formalization for this semantics.

\section{Conclusions and Future Work}

In this paper we presented a systematic extension of all major
semantics of logic programs to aggregate programs: the least fixpoint
semantics \cite{vanEmden76-JACM}, the standard model of stratified
programs \cite{Apt88-FDDLP}, the supported model semantics
\cite{Apt88-FDDLP}, the Kripke-Kleene semantics \cite{Fitting85-JLP},
the stable model semantics \cite{Gelfond88-ICSLP}, the well-founded
semantics \cite{VanGelder91-JACM}, and the ultimate stable and
ultimate well-founded semantics \cite{Denecker04-IC}. The extension of
the stable and well-founded semantics is not unique but is
parameterized by how aggregate relations are extended to three-valued
relations on three-valued sets. We studied three instances of these
semantics. Two of them are the least precise (called $triv$) and the
most precise (called $ult$) in this family and they are defined in a
uniform way for all aggregate relations. The third instance (called
$bnd$) is defined only for aggregate functions on totally ordered
sets. For some aggregates, most notably $\Sum$ and $\Prod$ the
$bnd$-semantics is strictly less precise than the $ult$ semantics.
Although, we did not present here a full complexity analysis, the advantage
of the $bnd$-semantics over the $ult$-semantics is that it has a lower
complexity \cite{Pelov04-PhD}.

We also showed that all important properties and relationships of the
original semantics are preserved in the extension. For example, a
stratified aggregate program $P$ has a two-valued $\ACD$-well-founded
model which is equal to the unique $\ACD$-stable model of $P$ and the
standard model of $P$ for any three-valued interpretation $\ACD$ of
the aggregate relations. A similar result for a definite aggregate
program $P$ is that the $ult$-well-founded model is equal to the unique
$ult$-stable model and the least fixpoint model of $P$. Another
important property of all stable semantics which we define is that
stable models are always minimal models.


\bibliographystyle{acmtrans}
\bibliography{aggr,asp,lp,logic,clp,complexity,math}
\appendix

\section{Proof of the Shortest Path Theorem}\label{app}

Before proving Proposition \ref{prop:sp-b}, we first introduce some concepts.

\newcommand{\shp}{sp}
\newcommand{\shpc}{spc}

We define the {\em length} of a path as its number of edges. We define
a partial function $\shp(.,.)$ as follows: $\shp(a,b)$ is defined if
and only if there is a shortest path from $a$ to $b$ and $\shp(a,b)=w$
where $w$ is the weight of the shortest path. Note that it is possible
that there is a path from $a$ to $b$ while there is not a shortest
path.  E.g., consider the graph with edges $\{(a,a,-1),(a,b,1)\}$. It
has no shortest paths.
 
\renewcommand{\theproposition}{\ref{prop:sp-b}}
\begin{proposition}
  Let $edge^\CD$ be a graph with the property that for any pair of
  nodes $a$ and $b$, if there is a path from $a$ to $b$, there is a
  shortest path from $a$ to $b$. In the $ult$-well-founded model of
  the shortest path program from Example~\ref{ex:sp-b} an atom
  $sp(a,b,w)$ is:
  \begin{itemize}
  \item {\em true} if a shortest path between $a$ and $b$ exists and
    has weight $w$;
  \item {\em false} otherwise.
  \end{itemize}
\end{proposition}
\begin{proof}
  We will compute the $ult$-well-founded model of $P$ by an alternating
  fixpoint computation using the sequences $(I_n)_{n\in \setN}$ and
  $(J_n)_{n\in \setN}$ which are defined by mutual induction:
  \begin{itemize}
  \item $I_0 = \bot$;
  \item $J_n = \ust{\Phi_P}(I_n) = \lfp(\Phi_P^2(I_n,\cdot))$;
  \item $I_n = \lst{\Phi_P}(J_{n-1}) = \lfp(\Phi_P^1(\cdot,J_{n-1}))$.
  \end{itemize}

  We now construct this sequence until we reach a fixpoint. 

  \begin{enumerate}
  \item We show that $J_0 = \lfp(\Phi_P^2(\bot,\cdot)) =  C\cup S$ where
    \begin{align*}
      C & = \{ cp(a,b,w) \vbar \text{there is a path from $a$ to $b$ with
        weight $w$}\} \\
      S & = \{ sp(a,b,w) \vbar \text{there is a path from $a$ to $b$ with
        weight $w$}\}.
    \end{align*}
    In the first iteration $\Phi_P^2(\bot,\bot) =
    T_{P,\CD}^{aggr}(\bot) = C_1$ where
    \[ C_1 = \{ cp(a,b,w) \vbar edge(a,b,w)\in edge^\CD \}. \]
    In the next iteration, the value of $cp/3$ will not change because
    the value of $sp/3$ has not changed. On the other hand $cp/3$ has
    changed, so $sp/3$ will change now. First, let us compute the value
    of the set expression for values $a$ and $b$:
    \[\eval{\se{c}{cp(a,b,c)}}{(\bot,C_1)} = (\emptyset,W_{ab})\]
    where $W_{a,b} = \{w\vbar (a,b,w)\in edge^\CD\}.$ i.e., there is a
    path between $a$ and $b$ of length 1.
    According to Table~\ref{tab:alg-extr} in
    Section~\ref{sec:aggr-ult}, it holds that
    $\ult{\Min}^2((\emptyset,W_{ab}),x)$ is true for all $x\in
    W_{ab}$. So, we obtain $\Phi_P^2(\bot,C_1) = C_1\cup S_1$ where
    \[ S_1 = \{ sp(a,b,w) \vbar edge(a,b,w)\in edge^\CD \}. \]
    Continuing this process we can show by induction that for every
    positive integer number $n>0$ we have
    \begin{align*}
      \Phi_P^2(\bot,\cdot)\uparrow(2n-1) &= \bigcup_{1\leq i \leq n}C_i \cup \bigcup_{1\leq i < n}S_i \\
      \Phi_P^2(\bot,\cdot)\uparrow(2n)   &= \bigcup_{1\leq i \leq n}C_i \cup \bigcup_{1\leq i \leq n}S_i
    \end{align*}
    where
    \begin{align*}
      C_i & = \{ cp(a,b,w) \vbar \text{there is a path from $a$ to $b$
        of length $i$ and weight $w$}\} \\
      S_i & = \{ sp(a,b,w) \vbar \text{there is a path from $a$ to $b$
        of length $i$ and weight $w$}\}.
    \end{align*}
    At the first limit ordinal $\omega$ we have 
    \[ \Phi_P^2(\bot,\cdot)\uparrow\omega = \bigcup_{i\in \setN}C_i\cup \bigcup_{i\in \setN}S_i = C \cup S. \]
    Hence, after this first step, we have computed in $J_0$ all possible
    path weights between any two pairs of points $a$ and $b$.

  \item Next we show that $I_1 = \lfp(\Phi_P^1(\cdot,C\cup S))=CE\cup SP$ where:
    \begin{align*}
      SP & = \{ sp(a,b,w) \vbar \text{there is a shortest path from $a$
        to $b$ of weight $w$}\} \\
      CE & =   \{ cp(a,b,w) \vbar 
      \begin{array}[t]{l}
        (a,b,w) \in edge^\CD    \text{ or } \\ 
        \exists c,w_1 \colon sp(a,c,w_1)\in SP  \text{ and  } (c,b,w-w_1)\in edge^\CD \}
      \end{array}
    \end{align*}

    Define also the following sets for every $i>1$: 
    \begin{align*}
      SP_i & = \{ sp(a,b,w) \vbar \text{there is a shortest path from $a$
        to $b$ of length $i$ and weight $w$ } \} \\
      CE_i & = \{ cp(a,b,w) \vbar
      \begin{array}[t]{l}
        (a,b,w) \in edge^\CD \text{ or } \\ 
                i>1 \text{ and } \exists c, w_1\colon sp(a,c,w_1)\in SP_{i-1}  \text{ and }
                (c,b,w-w_1)\in edge^\CD \}
      \end{array}
    \end{align*}
    Note that $SP = \bigcup_{i\in \setN} SP_i$ and $ CE = \bigcup_{i\in \setN} CE_i$. 

    For the first iteration we verify that
    \[ \Phi_P^1(\bot,C\cup S) = C_1 = CE_1. \]
    To see why this is the case, we compute  the value of the set
    expression: 
    $$\eval{\se{c}{cp(a,b,c)}}{(\bot,C\cup S)} = (\emptyset, A_{ab})$$
    where
    \[ A_{ab} = \{w\vbar \text{there is a path from $a$ to $b$ of
      length $w$} \}. \]

    Further, $\ult{\Min}^1((\emptyset,A_{ab}),w)$ is false for every weight
    $w$ (see Table~\ref{tab:alg-extr} of Section~\ref{sec:aggr-ult}).
    So, the interpretation of $sp/3$ will be the empty set. The
    interpretation of $cp/3$ will be the same as the $edge$ relation.
    Hence we obtain the set $CE_1$.

    On the next iteration only the value of $sp/3$ will change. We have
    \[ \eval{\se{c}{cp(a,b,c)}}{(CE_1,C\cup S)} = (W_{ab},A_{ab}) \]
    and $((W_{ab},A_{ab}),w)\in \ult{Min}^1$ if $w\in W_{ab}$ and the
    shortest path between $a$ and $b$ has weight $w$. So
    \[ \Phi_P^1(CE_1,C\cup S) = CE_1\cup SP_1.\]
    On the next iteration, the $cp/3$ relation becomes the composition of
    ${SP_1}$ with the $edge^\CD$ relation:
    \[ \Phi_P^1(CE_1\cup SP_1,C\cup S) = CE_1\cup CE_2\cup SP_1. \] In
    the next step we compute $CE_1\cup CE_2\cup SP_1\cup SP_2$, i.e.,
    we obtain all shortest paths of at most length 2. Continuing this
    process we obtain a fixpoint which is
    \[ \Phi_P^1(CE\cup SP,C\cup S) = CE\cup SP. \]
    At this stage, we have found all shortest paths. 
    
  \item Next, we show that $J_1=\lfp(\Phi_P^2(CE\cup SP,\cdot ))=CE\cup SP$,
    i.e., $\Phi_P^2(CE\cup SP,CE\cup SP)=CE\cup SP$.

    It is straightforward to verify that $cp(a,b,w) \in
    \Phi_P^2(CE\cup SP,CE\cup SP)$ if and only if $cp(a,b,w) \in CE$.
    As for  $sp(a,b,w)$, we first have to consider the value of the
    set expression 
    \[\eval{\se{c}{cp(a,b,c)}}{(CE\cup SP,CE\cup SP)} = (B_{ab},
    B_{ab})\] where $$B_{ab} = W_{ab}\cup\{ w \vbar \exists c, w_1\colon 
sp(a,c,w_1)\in SP \land (c,b,w-w_1) \in edge^\CD\}.$$ Either there is
no path from $a$ to $b$, in which case $B_{ab}$ is empty, and its
minimum undefined. In this case, no $sp(a,b,w)$ atom is derived. Or,
there is a path from $a$ to $b$. Then by the assumption of the
proposition, there is a shortest path from $a$ to $b$, say with weight
$w$. This minimal path is either an edge from $a$ to $b$ or it is an
extension of a shortest path from $a$ to some vertex $c$. In both
cases, $w$ belongs to the set $B_{ab}$ and is its least element. In
this case, $sp(a,b,w)$ is derived. Hence
\[\Phi_P^2(CE\cup SP,CE\cup SP) = CE\cup SP.\]

  \item Since $J_1=I_1$, it follows that $I_2=J_1=I_1$ and that we have
    reached a fixpoint which is the two-valued $ult$-well-founded
    model that was to be proven.

  \end{enumerate}
\end{proof}

The following example shows that the condition in the proposition is
essential for the proof.

\renewcommand{\theexample}{A.\arabic{example}}

\begin{example}
  Consider the following graph $\{ (a,a,-1), (a,b,0)\}$. It is easy to
  see that there are no shortest paths because of the cycle in
  $a$. Hence, $SP=\emptyset$.

  We compute the $ult$-well-founded model of $P$ by an alternating fixpoint
  computation. The first three steps are exactly as in the
  proof. Things only change in the computation of $J_1$. For this
  step, the proof exploited the fact that all connected pairs have a
  shortest path, but this assumption does not hold anymore. We have:

  \begin{itemize}
  \item $I_0 = \bot$.
  \item $J_0 = \{sp(a,a,-n-1),sp(a,b,-n),cp(a,a,-n-1),cp(a,b,-n)\vbar n\in
  \setN\}$. This describes all paths in the graph.
\item $I_1 = CE\cup SP = \{cp(a,a,-1),cp(a,b,0)\}$. Indeed, since
  there are no shortest paths in this graph, the set
  $SP$ is empty and $CE$ is just a
  copy of the edge relation.

  \item $J_1=\lfp(\Phi_P^2(CE,\cdot ))$. The computation is entirely
  similar to the fixpoint computation of $J_0$ (see the proof of the
  proposition). Define $$C_i = \{cp(a,a,-n-1),cp(a,b,-n)\vbar n\in[0,i]\}$$
  $$S_i = \{sp(a,a,-n),sp(a,b,-n+1)\vbar n\in[0,i[\}.$$ Note that $CE =
  C_0 \cup S_0$. First, we compute $\Phi_P^2(CE,CE)$. Since there are
  no true $sp$ atoms, the computed $cp$ atoms correspond to the
  edges. Hence, $cp$ remains unchanged. As for $sp$, its rule derives
  the atoms $sp(a,a,-1)$ and $sp(a,b,0)$, i.e., we obtain $C_0\cup
  S_1$. In the next iteration, since $cp$ did not change, $sp$
  remains unaltered. Now $cp$ is extended by composing $S_1$ with
  the edge relation. We obtain $C_1$. In the next iteration, $cp$ 
  remain identical, and now $sp$ will be extended to obtain
  $S_2$.  In general we have the same fixpoint computation as for
  $J_0$ except that we start at $CE$ rather than at $\bot$.  It holds
  that 
  \begin{align*}
    \Phi_P^2(CE,\cdot)\uparrow(2n) &= C_n \cup S_n \\
    \Phi_P^2(CE,\cdot)\uparrow(2n+1)   &= C_n \cup S_{n+1}
  \end{align*}
  The limit of this sequence, $J_1$, is equal to $J_0$. Therefore,
  $I_2$ will be equal to $I_1$, so we reached the well-founded
  fixpoint which is $(I_1,J_0)$.  Since $I_1 \neq J_0$, this is a
  three-valued model. For example, for each negative integer $n$, the
  atom $sp(a,b,n)$ is unknown. \qed
  \end{itemize}
\end{example}

\end{document}